%% file: main.tex
\documentclass[
    reprint,
    aps,
    prl,
	amsmath,
	amssymb,
	superscriptaddress,
	longbibliography,
	english
]{revtex4-2}

\usepackage{graphicx}
\usepackage{physics}
\usepackage[dvipsnames]{xcolor}
\usepackage[hidelinks,colorlinks=True,linkcolor=BlueViolet,filecolor=BlueViolet,urlcolor=BlueViolet,citecolor=BlueViolet]{hyperref}
\usepackage[all]{hypcap}

\begin{document}
\title{Effective non-local parity-dependent couplings in qubit chains}

\author{Maximilian N\"{a}gele}
\affiliation{Fakult\"{a}t f\"{u}r Physik, Ludwig-Maximilians-Universit\"{a}t M\"{u}nchen, Schellingstraße 4, D-80799 M\"{u}nchen, Germany}
\affiliation{Walther-Mei{\ss}ner-Institut, Bayerische Akademie der Wissenschaften, 85748 Garching, Germany}

\author{Christian Schweizer}
\affiliation{Fakult\"{a}t f\"{u}r Physik, Ludwig-Maximilians-Universit\"{a}t M\"{u}nchen, Schellingstraße 4, D-80799 M\"{u}nchen, Germany}
\affiliation{Walther-Mei{\ss}ner-Institut, Bayerische Akademie der Wissenschaften, 85748 Garching, Germany}
\affiliation{Munich Center for Quantum Science and Technology (MCQST), Schellingstra\ss e 4, 80799 M\"{u}nchen, Germany}

\author{Federico Roy}
\affiliation{Walther-Mei{\ss}ner-Institut, Bayerische Akademie der Wissenschaften, 85748 Garching, Germany}
\affiliation{Theoretical Physics, Saarland University, 66123 Saarbr\"ucken, Germany}

\author{Stefan Filipp}
\affiliation{Walther-Mei{\ss}ner-Institut, Bayerische Akademie der Wissenschaften, 85748 Garching, Germany}
\affiliation{Munich Center for Quantum Science and Technology (MCQST), Schellingstra\ss e 4, 80799 M\"{u}nchen, Germany}
\affiliation{Physik-Department, Technische Universit\"{a}t M\"{u}nchen, 85748 Garching, Germany}

\begin{abstract}
For the efficient implementation of quantum algorithms, practical ways to generate many-body entanglement are a basic requirement.
Specifically, coupling multiple qubit pairs at once can be advantageous and can lead to multi-qubit operations useful in the construction of hardware-tailored algorithms.
Here we harness the simultaneous coupling of qubits on a chain and engineer a set of non-local parity-dependent quantum operations suitable for a wide range of applications.
The resulting effective long-range couplings directly implement a parametrizable Trotter-step for Jordan-Wigner fermions and can be used for simulations of quantum dynamics, efficient state generation in variational quantum eigensolvers, parity measurements for error-correction schemes, and the generation of efficient multi-qubit gates.
Moreover, we present numerical simulations of the gate operation in a superconducting quantum circuit architecture, which show a high gate fidelity of $>99.9\%$ for realistic experimental parameters.
\end{abstract}

\date{March 14, 2022}
\maketitle
In recent years, the field of quantum computing has made significant advances in demonstrating applications where quantum devices are predicted to be advantageous \cite{Kim2021, Arute2020, Nam2020, Hempel2018}.
A promising near-term application is the simulation of quantum mechanical systems~\cite{Cirac2012, Bharti2021}. In particular, the simulation of fermionic systems is important to predict the properties of e.g.\ molecules~\cite{Moll2018, Mcardle2020}, or to understand many-body systems such as the Fermi-Hubbard model, which is expected to explain phenomena of great scientific and industrial interest like high-temperature superconductivity~\cite{Esslinger2010}. However, mapping fermions to qubits poses a major challenge, since local fermionic couplings can result in non-local qubit interactions~\cite{Havl2017, Bravyi2002, Nielsen2005}.

To build quantum processors, different physical platforms, such as trapped ions~\cite{Bruzewicz2019}, superconducting qubits~\cite{Clarke2008, Krantz2019}, quantum dots~\cite{Kloeffel2013}, neutral atoms~\cite{Henriet2020} and photonic qubits~\cite{Kok2007} are currently considered.
Independent of the platform, an important characteristic for each device's capability is the qubit connectivity, which is typically limited to local two-body couplings~\cite{Linke2017}, while
non-local interactions are challenging to implement and require a large amount of consecutive two-qubit gates~\cite{Cowtan2019} or
ancilla qubits~\cite{Kempe2005, Babbush2013}.

An alternative solution is to implement non-local terms by controlling multiple two-body couplings simultaneously~\cite{Zhang2021, Kranzl2021, Gu2021, Burkhart2021, Glaser2022}. A prime example of such a method is the perfect state transfer along a qubit chain~\cite{Christandl2004, Yung2006, Vinet2012, Chapman2016, Brougham2011, Nikolopoulos2004, Shi2005, Albanese2004, Vinet2012_Krawtchouk}, where 
an excitation at an initial location is transferred to a final location along the chain. This technique has a large variety of applications, such as entanglement generation and effective two-qubit gates~\cite{Yung2005, Kay2010, Li2018, Nielsen2002, Yung2004}. Recently, it has been extended to fractional state transfer (FST)~\cite{Genest2016, Chan2019, Lemay2016}, where the quantum state is partially transferred to the final location,
while the other part returns to its original position.

\begin{figure}[h!]
    \centering
    \includegraphics[width=\linewidth]{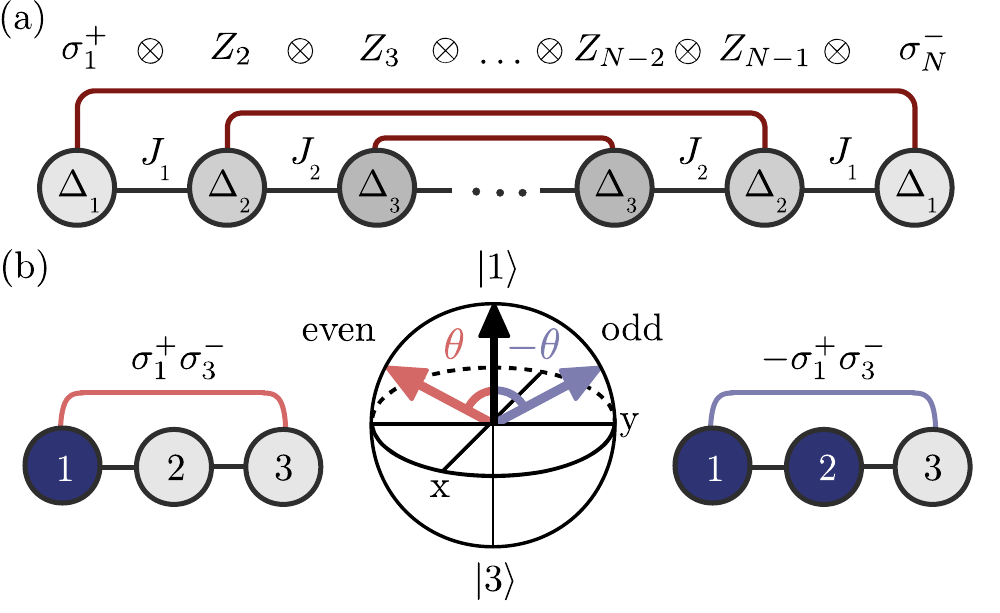}
    \caption{Qubit chain and effective parity-dependent couplings.
    (a) Chain of qubits (circles) with frequencies $\Delta_n$ and direct couplings $J_n$, as described by $H_N$ in Eq.~\eqref{eq:H_chain}.
    Dark red lines indicate effective non-local interactions that stroboscopically arise for specific parameter choices of $\Delta_n$ and $J_n$.
    The effective interaction results in a rotation in the subspaces spanned by $\ket{n}$ and $\ket{N+1-n}$, where $n$ denotes the location of the excitation.
    The orientation of the rotation vector depends on the parity of the qubits between each pair, $\otimes_{k=n+1}^{N-n} Z_k$.
    (b) Illustration of a chain with length $N=3$.
    A single excitation is prepared at site $1$ and partially transferred to site 3 with effective interaction $\sigma^+_1 \sigma_3^- + \text{h.c.}$ (left chain), which rotates the state by an angle $\theta$ on the Bloch-sphere spanned by the states $\ket{1}$ and $\ket{3}$ (red arrow).
    If an additional excitation is prepared at site $2$ (right chain), the effective interaction changes sign, such that the state rotates by an angle $-\theta$ on the Bloch-sphere (blue arrow).
} \label{fig:chain_effective}
\end{figure}

In this work, we build on FST and harness nearest-neighbor couplings in a linear chain of two-level systems to engineer effective non-local interactions that depend explicitly on the number of excitations in the chain (see Fig.~\ref{fig:chain_effective}).
These interactions directly implement fermionic couplings between qubits on opposite sides of the chain under Jordan-Wigner transformation (JWT) and, thus, generate a set of matchgates~\cite{Valiant2002}, which correspond to unitary evolution of free fermions~\cite{Terhal2002, Knill2001}.
In addition to fermionic quantum simulation, the excitation-dependent operation also provides an efficient way to measure long strings of qubit correlators with potential applications in quantum error correction~\cite{Cohen2021,Lidar2013}.

\begin{figure}[t]
    \includegraphics[width=\linewidth]{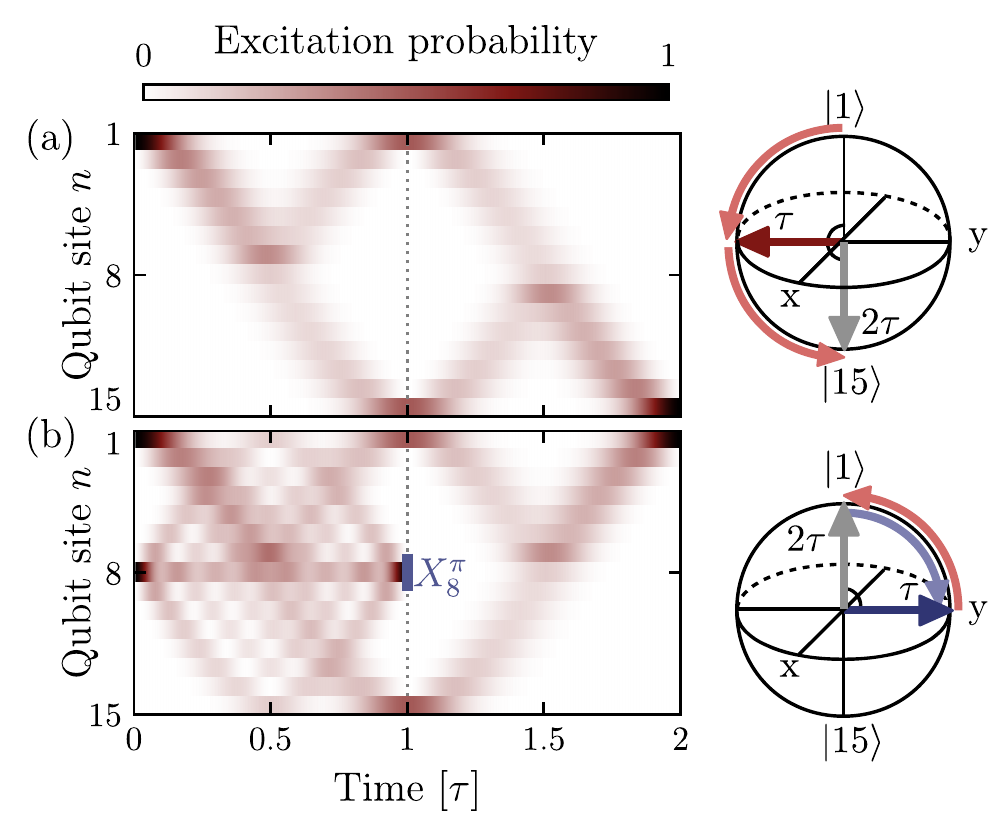}
    \caption{
    Simulated occupation dynamics
    during two consecutive fractional state transfers (FST) on a $N=15$ qubit chain. Hamiltonian parameters are chosen to achieve a transfer angle $\theta=\pi/2$.
    (a)
    Evolution of an excitation prepared at site~$1$. After a first
    FST at time $\tau$, (dotted line) the excitation is in a superposition of
    both ends of the chain. In the Bloch-sphere spanned by the states $\ket{1}$ and $\ket{15}$ this corresponds to a rotation by an angle~$\pi/2$ (dark red arrow).
    At time $2\tau$ after a second FST, the excitation refocuses at site~$15$, resulting in a $2\theta = \pi$ rotation on the Bloch-sphere (gray arrow).
    (b) Evolution of a state with excitations prepared at sites~$1$ and $8$. 
    In this case, the first FST rotates the state $\ket{1}$  on the Bloch-sphere by a negative angle -$\pi/2$ (dark blue arrow), due to the odd parity of excitations in the middle of the chain.
    The center excitation instead refocuses on its original location and is then removed by an instantaneous $\pi$-flip gate $X_8^\pi$ at site 8 (blue square), which changes the parity.
    Then, with a second FST the dynamics are reverted and the excitation refocuses at site~$1$.
    }\label{fig:dynamics_comp_phi0.5_reflection}
\end{figure}

The Hamiltonian of a qubit chain with length $N$ is given by
\begin{equation} \label{eq:H_chain}
 H_N  = \sum_{n=1}^N \Delta_n \sigma_n^+ \sigma_n^- + \sum_{n=1}^{N-1} \left(J_n \sigma^+_n \sigma_{n+1}^-+ \text{h.c.}\right),
\end{equation}
where we assume $\hbar=1$. Here, $\sigma^\mp_n$ are the qubit lowering (raising) operators and $\Delta_n$ is the frequency of qubit~$n$.
The coupling between qubits $n$ and $n+1$ is mediated via $XY$-interactions $\sigma^+_n \sigma^-_{n+1} + \text{h.c.} = (X_n X_{n+1} + Y_n Y_{n+1}) / 2$ with time-independent coupling strengths $J_n$. 
We use the notation $X,Y,Z$ and $I$ for the Pauli matrices and the identity.

To implement FST, we set $J_n = J_{N-n}$ and $\Delta_n = \Delta_{N+1-n}$ to be symmetric about the center.
Since the Hamiltonian commutes with the total excitation number operator, the total number of excitations in the system is preserved and each excitation manifold can be considered separately.
Hence, we first consider FST in the single-excitation manifold, where $H_N$ is tridiagonal and persymmetric, i.e.\ symmetric around its antidiagonal:
\begin{equation}
H_N^{(1)} = 
\left( \begin{array}{rrrrrr}
\Delta_1 & J_1 &  & & \\ 
J_1 & \Delta_2 & J_2 & & \\ 
&\ddots & \ddots  &\ddots & \\
 & & J_2 & \Delta_2 & J_1\\
 & &  & J_1 & \Delta_1\\
\end{array} \right).
\end{equation}
As such, it has only mirror-symmetric and mirror-antisymmetric eigenvectors, $\ket{v_n^s}$ and $\ket{v_n^a}$, with real non-degenerate eigenvalues $\lambda_n^{s/a}$~\cite{Hochstadt1967,Kay2010}. Hence, we can expand the single-excitation basis state $\ket{n} = \ket{0...1_n...0}$ and its mirror state $\ket{N+1-n}$ as
\begin{align} 
\begin{aligned}
\ket{n} &= \sum_n \alpha_n^s \ket{v_n^s} + \sum_n  \alpha_n^a \ket{v_n^a},
\\
\ket{N+1-n} &= \sum_n \alpha_n^s \ket{v_n^s} - \sum_n  \alpha_n^a \ket{v_n^a}.
\label{eq:expand_n_center_sym}
\end{aligned}
\end{align}
Specific transfer angles~$\theta$ between mirror-symmetric states can be achieved by choosing the parameters $J_n$ and $\Delta_n$ such that the eigenvalues of $H_N^{(1)}$ have the form
\begin{equation} \label{eq:lambda_s/a}
\lambda_n^{s/a}\tau = \pm\frac{\theta}{2} + \phi + m_n^{s/a} 2\pi,
\end{equation}
with $m_n^s,\,m_n^a\in\mathbb{Z}$ and $\phi$ 
a phase acquired during transfer (see \hyperref[sup:Hparas]{Supplemental Material}). 
Evolving the state~$\ket{n}$ according to these eigenvalues and eigenstates
for transfer time $\tau$ results in
\begin{equation} 
\begin{aligned}
\label{eq:ev_sym_as}
e^{-i H_N^{(1)} \tau} \ket{n} 
= \sum_n e^{-i \lambda_n^s \tau} \alpha_n^s \ket{v_n^s} + \sum_n e^{-i \lambda_n^a \tau} \alpha_n^a \ket{v_n^a} \\
=e^{-i \phi} \left(
\cos(\frac{\theta}{2})\ket{n} - i \sin(\frac{\theta}{2})\ket{N+1-n} \right).
\end{aligned}
\end{equation}
Thus, qubit $n$ and its mirror qubit on the chain, qubit $N +1-n$, are rotated by an angle $\theta$ in their respective two-qubit subspace, which realizes FST.
We simulate the dynamics in the single-excitation manifold for a chain with length $N=15$ and parameters such that $\theta=\pi/2$ as shown in Fig.~\ref{fig:dynamics_comp_phi0.5_reflection}(a).
After time~$\tau$, a system initially in $\ket{1}$ is rotated to the superposition state $(\ket{1} - i \ket{15})/\sqrt{2}$. After time $2\tau$, the total transfer angle is $2 \theta = \pi$ and the excitation refocuses in state $\ket{15}$.

\textit{Parity-dependent rotations.}\,---\,%
We now consider the full unitary evolution under $H_N$ for arbitrary initial states. We provide an intuition for the resulting interactions between qubits by mapping the time-evolution of $H_N$ to the dynamics of an effective non-local Hamiltonian
\begin{equation}
\begin{split}
G_N =\,& \sigma^+_1\otimes Z_2 \otimes Z_3 \otimes \dots{} \otimes Z_{N-2}\otimes Z_{N-1} \otimes \sigma^-_N
\\
+&\,\,I_1 \,\otimes \sigma^+_2\otimes Z_3 \otimes \dots{}\otimes Z_{N-2} \otimes \sigma^-_{N-1} \otimes I_N
\\
+& \dots{} + \mathrm{h.c.}\,.
\end{split}\label{eq:GN}
\end{equation}
At integer multiples of the transfer time $\tau$ the
time evolution under $H_N$ generates the same unitary, up to single-qubit phases, as the evolution under $G_N$ for a transfer angle $\theta$.
Indeed, the unitary $K_N = \exp(-i \theta/2 G_N )$ can be realized by FST through
\begin{equation}
\exp(-i \frac{\theta}{2} G_N ) = \exp(-i H_N \tau) \exp(i \phi H_{\text z}),
\label{eq:mapping}
\end{equation}
where \mbox{$H_{\text z} = \sum_n \sigma_n^+ \sigma_n^-$} accounts for phase difference by local unitary transformation and would include an additional phase $\theta/2$ for the middle qubit in odd chains.

The form of $G_N$ explicitly shows the parity-dependent mirror-symmetric rotation of excitations along the chain.
Therefore, at stroboscopic times, the evolution under $H_N$ can be understood as a rotation between each pair of mirror qubits, where the sign of the rotation angle is given by the parity of all qubits between them (see Fig.~\ref{fig:chain_effective} and~\ref{fig:dynamics_comp_phi0.5_reflection}). 
Since the different terms in the sum of Eq.~\eqref{eq:GN} commute, these rotations are independent of each other.

To prove Eq.~\eqref{eq:mapping}, we analyze the Hamiltonians $H_N$, $G_N$ and $H_\text z$ in terms of fermionic operators using a Jordan-Wigner transformation (JWT)~\cite{Nielsen2005}.
We find that all transformed Hamiltonians
describe non-interacting fermions. Therefore, their complete dynamics can be constructed from the single-excitation manifold using Slater-determinants~\cite{Cappellaro2011}.
Since the single-excitation dynamics of both sides of the equation are equivalent, this construction leads to the same unitary evolution and, therefore, Eq.~\eqref{eq:mapping} holds in all excitation manifolds 
(see \hyperref[sup:mappGH]{Supplemental Material}).

To demonstrate the parity-dependence, we simulate the time evolution under $H_N$ for two consecutive FST processes with $\theta = \pi / 2$ and with a parity change between them.
We prepare two excitations, one at the origin and one
at the center of the chain. After evolving for time $\tau$, the excitation from the origin of the chain is partially transferred to the other end of the chain, while the excitation in the middle of the chain refocuses at the same site.
We then remove the center excitation with an instantaneous $X$ gate, thus changing the parity in the center of the chain. Evolving for a further time $\tau$, the rotation angle $\theta$ is now inverted, causing a reversal of the dynamics as shown in Fig.~\ref{fig:dynamics_comp_phi0.5_reflection}(b). Indeed, the initial excitation at site~$1$ returns to its original position, in contrast to the dynamics of Fig.~\ref{fig:dynamics_comp_phi0.5_reflection}(a) where the parity is identical for both FST processes.

\begin{figure}[t]
    \includegraphics[width=0.99\linewidth]{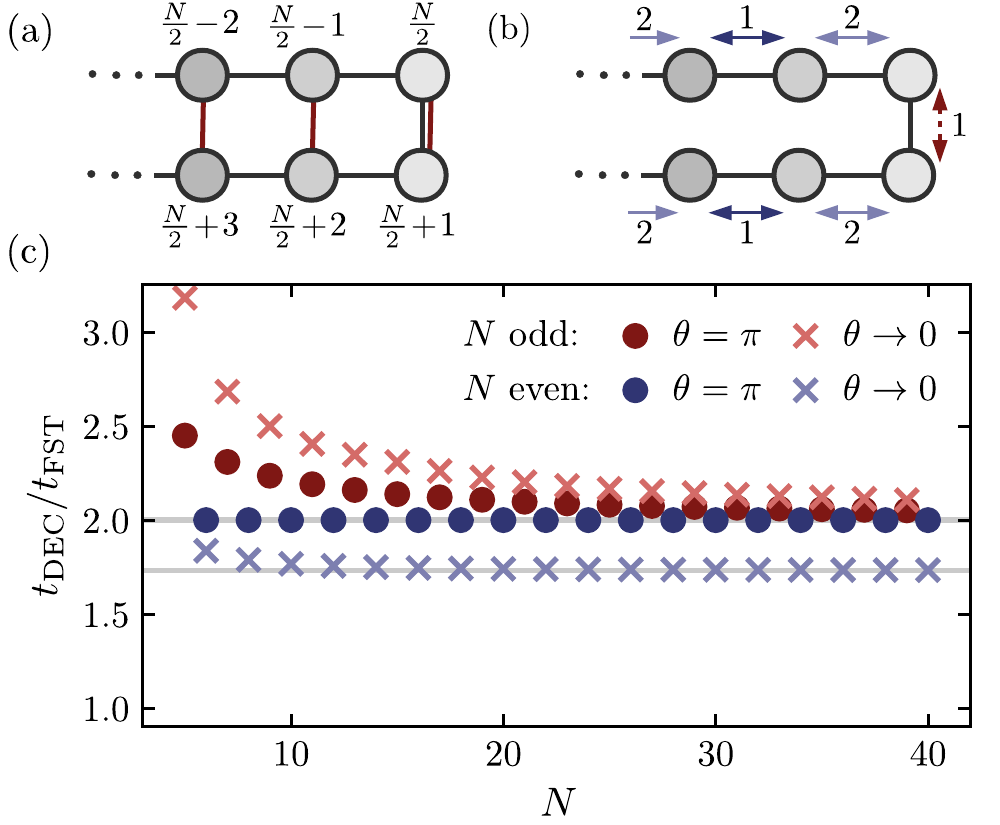}
    \caption{
    Comparison between fractional state transfer (FST) and equivalent decomposition. (a) A
    chain folded in the middle forms a ladder, where FST  introduces effective interactions along its rungs corresponding to the different terms of $G_N$ (red lines). (b) Decomposition of FST into two-qubit gates for even chains. Step one (red and dark blue arrows) and two (light blue arrows) are repeated $N/2$ times. Blue arrows symbolize FSWAP gates while the red arrow is an iSWAP$(\theta)$ gate. Odd chains are discussed in the \hyperref[sup:Decomp]{Supplemental Material}. (c) 
    Relative speed gain of FST with respect to the two-qubit gate decomposition of $\exp(-i\theta G_N/2)$ as a function of the chain length $N$, for both odd (red) and even (blue) $N$.
    For perfect state transfer, i.e.\ $\theta = \pi$, FST is at least a factor of two (upper solid line) faster. For transfer angles  approaching zero, i.e.\ $\theta \to 0$, the same speed-up is still present for $N$ odd but reduces to $\sqrt{3}$ (lower solid line) for $N$ even.
    }\label{fig:time_comp}
\end{figure}

\textit{Applications}\,---\,%
Fermions can be simulated on a quantum computer by using the Jordan-Wigner transformation~\cite{Ortiz2001, Cade2020}
with fermionic annihilation operators
\mbox{$a_n = -(\otimes_{k=1}^{n-1} Z_k) \otimes \sigma^-_n$}.
For one-dimensional fermionic systems nearest-neighbor couplings are easily simulated on qubit systems with local two-qubit gates~\cite{Lieb1961}. 
However, in two-dimensional systems or ladder-type geometries, nearest-neighbor couplings are challenging because the one-dimensional structure of the Jordan-Wigner encoding leads to non-local operators.
The JWT of $G_N$, 
\mbox{$G_N^{\text{F}} = a_1^\dagger a_N +  a_2^\dagger a_{N-1}+ \dots{} + \text{h.c.}$},
creates long-range couplings between distant fermion sites, which can be used to implement such non-local terms. For example, when folding an even chain in half, all rung couplings of the system are directly implemented by $G_N$, which enables efficient simulation of the fermionic dynamics. To assess the efficiency, we implement the evolution under $G_N$ for arbitrary times by either applying FST with the correct rotation angle or by decomposing its action with consecutive two-qubit gates. Assuming that the gate speed is limited by the maximum achievable coupling $J_{\text{max}}$ we find that by applying FST we can achieve a speed-up of at least two-fold for odd $N$ and $\sqrt{3}$ for even $N>4$, with greater improvements for shorter chain lengths
(see Fig.~\ref{fig:time_comp}).
The non-local couplings of FST can, therefore, be used to implement fast Trotter simulations of fermionic systems.

Due to their native parity-dependent property, FST gates can also be harnessed to quickly measure correlators on long qubit chains, with applications in error correction e.g.\ in low-density parity-check codes~\cite{Cohen2021}.
To this end, consider a qubit chain in the state $\ket{\psi_m}$ with $m$ excitations. Its parity can be measured by introducing an ancilla qubit on each end of the chain. After applying the sequence of gates
\begin{equation}
\begin{aligned}
\left[X_{\text L}^{\frac{\pi}{2}} K_{N+2}^{(\pi)} Y_{\text R}^{\frac{\pi}{2}}\right]& \ket{0}_\text L\ket{\psi_m}\ket{0}_\text R
\\&\hspace{-5mm}= 
\begin{cases}
- i \ket{1}_\text L\ket{\bar\psi_m}\ket{0}_\text R \text{ for } m \text{ even}
\\
\hphantom{-i} \ket{0}_\text L\ket{\bar\psi_m}\ket{0}_\text R \text{ for } m \text{ odd}
\end{cases}
\end{aligned}
\end{equation}
a measurement of the left ancilla reveals the parity of excitations in the qubit chain. Here, $\ket{\bar\psi_m} = K_{N}^{(\pi)} \ket{\psi_m}$,  $X_{\text L}^{\frac{\pi}{2}}$ ($Y_{\text R}^{\frac{\pi}{2}}$) is a $\pi$-half $X$ ($Y$)-rotation on the left (right) ancilla and $K_{N+2}^{(\pi)}$ is the FST gate on the extended chain including the ancillas with $\theta = \pi$. This protocol requires $t \approx \frac{N+2}{2} \tau_\text{iSWAP}$, where $\tau_\text{iSWAP} = \frac{\pi}{2 J_{\text{max}}}$ is the time required for a nearest neighbor iSWAP gate. In comparison, a protocol based on two-qubit gates would require at least $N$ such gates. 
The operation on the middle part of the chain can be reversed by applying $K_{N}^{(\pi)}$, increasing the required time to $t \approx (N + 1) \tau_\text{iSWAP}$.
However, for applications that perform repeated parity measurements of the same chain, this reversion is unnecessary since $(K_N^{(\pi)})^2 = I$ up to single-qubit $Z$-rotations.
Furthermore, by applying single-qubit rotations before and after the measurement to introduce a basis change, any desired combination of correlators $P_1 P_2 \dots{} P_N$ with $P\in \{X,Y,Z\}$ can be measured.

\begin{figure}
    \includegraphics[width=\linewidth]{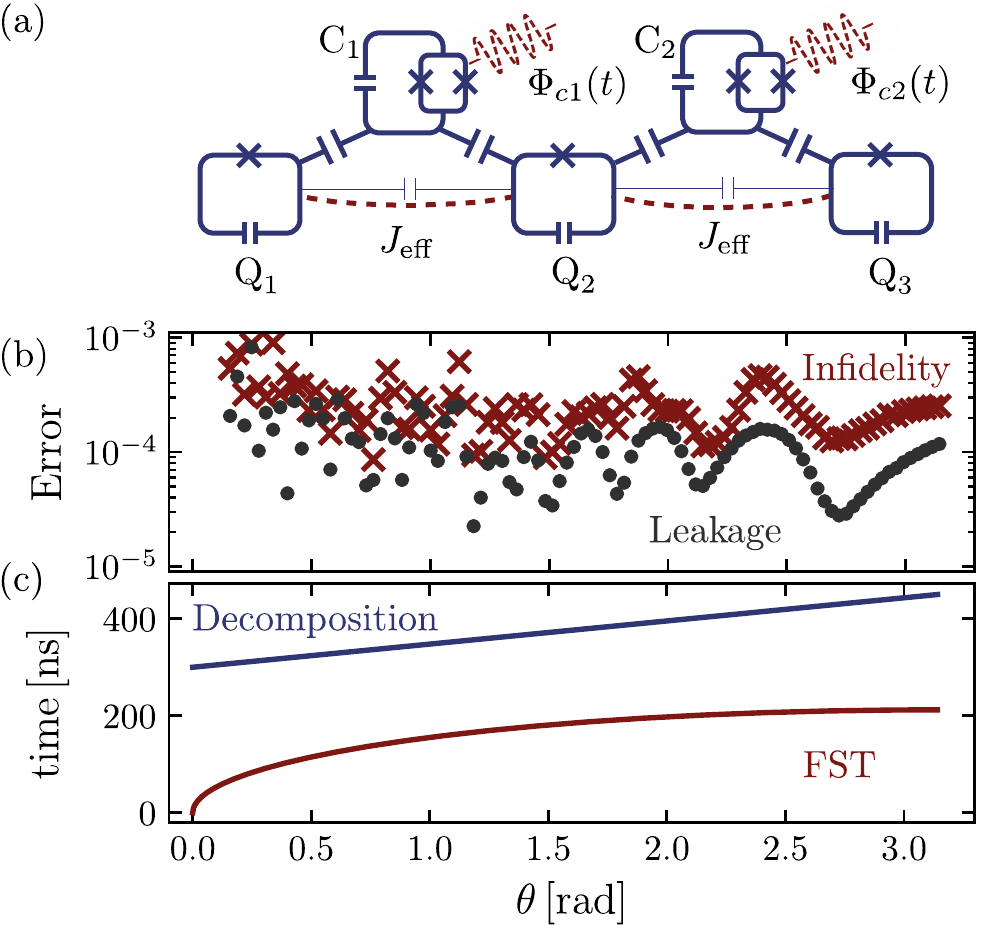}
    \caption{
    Simulation of the three-qubit fractional-state-transfer gate for superconducting qubits.
    (a)
    Three fixed-frequency transmons (Q$_1$, Q$_2$, Q$_3$) are coupled via two tunable couplers (C$_1$, C$_2$). By periodically modulating the coupler frequencies through flux pulses ($\Phi_{c1}$, $\Phi_{c2}$) effective interactions between the qubits arise ($J_{\text{eff}}$) 
    (b) Infidelity and leakage (average population loss after the gate per computational state) of the optimized gate at various angles.
    (c) The total gate time for different $\theta$ is set as theoretically predicted (red). The gate time is significantly shorter than the decomposition into two-qubit gates (blue).
    }\label{fig:sim_res}
\end{figure}

Applying FST on a three-qubit chain leads to an interesting multi-qubit gate, which directly implements parametrizable fermionic next-nearest neighbor interaction under JWT, 
\begin{equation*}
    K_3 = \ket{0}\bra{0}_2 \otimes \text{iSWAP}_{13}(- \theta) +  \ket{1}\bra{1}_2 \otimes \text{iSWAP}_{13}(\theta),
\end{equation*}
where the indices indicate the qubit positions.
We assume that the gate speed is limited by the maximum coupling $J_\text{max}$, since high detunings are usually experimentally feasible~\cite{Borjans2020,Koch2007,Hutchings2017}. Given this assumption, the FST
gate is significantly faster than its decomposition in two-qubit gates given by $\text{FSWAP}_{12}\text{iSWAP}_{23}(-\theta)\text{FSWAP}_{12}$, where FSWAP is the fermionic swap gate~\cite{Cade2020} [Fig.~\ref{fig:sim_res}(c)]. Since this speed-up increases for smaller angles, the FST gate is well suited for variational quantum algorithms and Trotter simulations, which often require only short interaction steps~\cite{Mcclean2016}.

To assess the experimental feasibility of this three-qubit gate, we numerically simulate it in a superconducting architecture, using the q-optimize software package~\cite{Wittler2021}. The simulated setup contains three fixed-frequency transmons that are dispersively coupled with two flux tunable coupler transmons as shown in Fig.~\ref{fig:sim_res}(a). By periodically modulating the frequency of a coupler via a flux drive, an effective coupling of its neighboring qubits can be realized~\cite{Mckay2016, Sete2021} (see \hyperref[sup:Eff_H_der]{Supplemental Material} for more details on the simulation and the chosen parameters).
We optimize the gate for a range of effective transfer angles $\theta \in [0.05\pi, \pi]$ assuming perfect single-qubit virtual $Z$ gates~\cite{Mckay2017, Zhu1997}.
We show that, without taking finite coherence into account, average infidelities~\cite{NielsenFidelity2002} lower than $10^{-3}$ are achieved for the whole range of the transfer angle $\theta$  [see Fig.~\ref{fig:sim_res}(b)]. 
The fidelity is mainly limited by leakage, which oscillates periodically with the gate length. This leakage is caused by off-resonantly driving transitions to the second-excited qubit states via higher harmonics of the drive and could be mitigated by pulse shaping~\cite{Motzoi2009, Werninghaus2021} or engineering qubits with higher anharmonicities~\cite{Yan2020}.

\textit{Conclusion.}\,---\,%
We have demonstrated how simultaneous nearest-neighbor couplings between qubits on a chain can be harnessed to generate dynamics equivalent to complex non-local interactions.
Building on FST, we have engineered long-range couplings dependent on qubit correlators along the chain. These directly implement fermionic coupling terms under a Jordan-Wigner transformation.
The resulting multi-qubit gates provide a significant speed-up compared to an equivalent decomposition into two-qubit gates, making them promising candidates for the implementation of fermionic simulation or as a building block in quantum variational algorithms.
Furthermore, we have shown that the parity-dependent property of FST gates can be harnessed for efficient measurements of qubit correlators, with applications in quantum error correction or quantum phase recognition~\cite{Herrmann2021}. We performed realistic numerical simulations of a superconducting-circuit three-qubit chain suggesting gate fidelities above $99.9\%$ under coherent evolution.
In the next step, we can extend the current protocol by introducing time-dependent controls. Since qubit chains can be fully understood in the single-excitation manifold, even for large systems numerical simulations remain tractable. Therefore, optimal-control techniques can be used to explore the space of possible operations, thus enabling the discovery of a variety of high-fidelity multi-qubit gates.

\textit{Acknowledgements.}\,---\,%
We thank Cosimo Rusconi, Ignacio Cirac, and Monika Aidelsburger for insightful discussions. 
F.R. acknowledges funding by the European Commission Marie Curie ETN project QuSCo (Grant Nr. 765267) and by GeCQoS (Grant Nr. 13N15680) project funded by the Federal Ministry of Education and Research (BMBF).
C.S. has received funding from the European Union’s Framework Programme for Research and Innovation Horizon 2020 (2014-2020) under the Marie Sk{\l}odowska-Curie Grant Agreement No.\,754388 (LMUResearchFellows) and from LMUexcellent,
funded by the BMBF and the Free State of Bavaria under the
Excellence Strategy of the German Federal Government and the L\"ander.
We also acknowledge funding by the Deutsche Forschungsgemeinschaft (DFG, German Research Foundation) – project number FI 2549/1-1.
\input{main.bbl}

%%%%%%%%%% Merge with supplemental materials %%%%%%%%%%
\clearpage
\widetext
\begin{center}
\textbf{\large Effective non-local parity-dependent couplings in qubit chains: Supplemental Material}
\end{center}
\twocolumngrid
%%%%%%%%%% Merge with supplemental materials %%%%%%%%%%
%%%%%%%%%% Prefix a "S" to all equations, figures, tables and reset the counter %%%%%%%%%%
\setcounter{equation}{0}
\setcounter{figure}{0}
\setcounter{table}{0}
\setcounter{page}{1}
\makeatletter
\renewcommand{\theequation}{S\arabic{equation}}
\renewcommand{\thetable}{S\arabic{table}}

\renewcommand{\thefigure}{S\arabic{figure}}
\renewcommand{\bibnumfmt}[1]{[S#1]}
\renewcommand{\citenumfont}[1]{S#1}

\renewcommand{\theHtable}{S\arabic{table}}
\renewcommand{\theHequation}{S\arabic{equation}}
\renewcommand{\theHfigure}{S\arabic{figure}}
%%%%%%%%%% Prefix a "S" to all equations, figures, tables and reset the counter %%%%%%%%%%

\section{S-I. Hamiltonian Parameters for FST} \label{sup:Hparas}
\begin{figure*}[t]
    \includegraphics[width=0.9\linewidth]{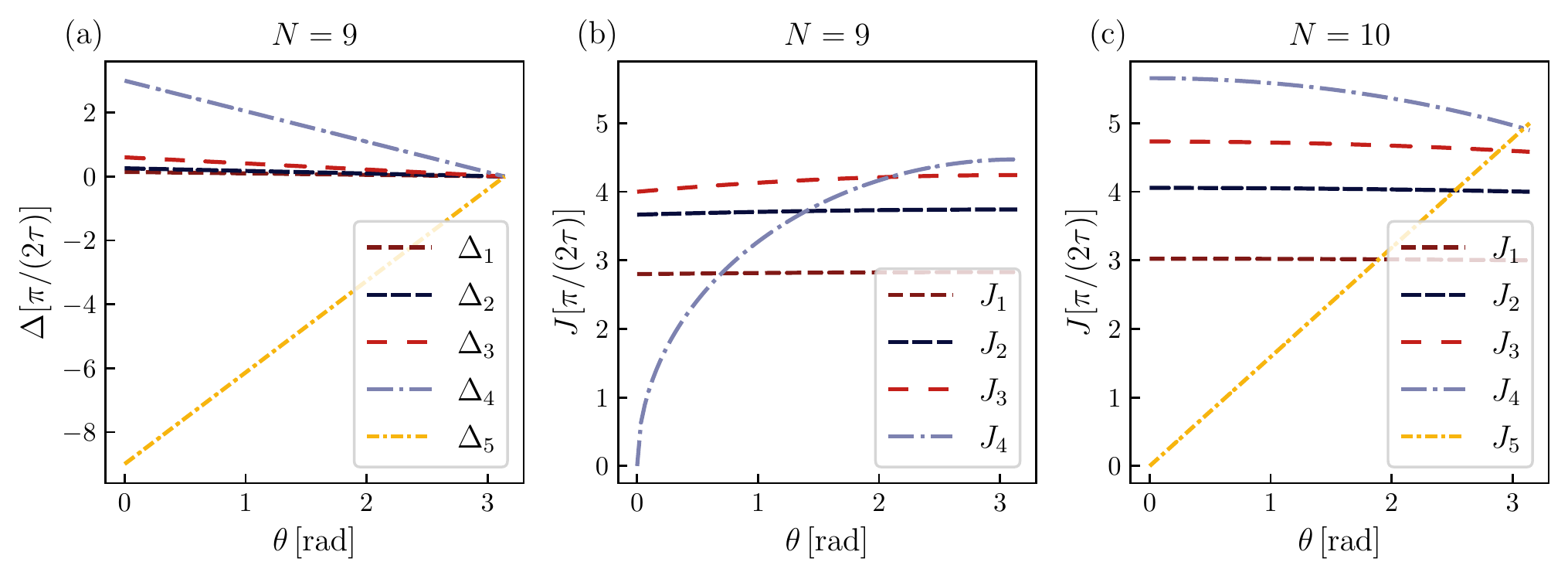}
    \caption{Hamiltonian Parameters $J_n$ and $\Delta_n$ as function of the transfer angle $\theta$ for chain length $N=9$ in (a) and (b) and $N=10$ in (c).
    (a) The magnitude of the required detunings decreases linearly with $\theta$ if $N$ odd. Perfect state transfer requires no detunings. For $N$ even no detunings are needed. (b) All coupling strengths increase with $\theta$  if $N$ odd. (c) For $N$ even only the center coupling increases with $\theta$ while all other couplings decrease. \label{fig:JD}}
\end{figure*}

In this supplement we use the notation defined in the main text. Given its eigenvalues a mirror-symmetric tridiagonal matrix can be uniquely reconstructed. Taking the spectrum of $H_N^{(1)}$ to be as narrow as possible ($m_n^s = m_n^a$, $m_{n+1}^{s/a} = m_n^{s/a} + 1$) and the gate time to be $\tau$, the required Hamiltonian parameters to achieve FST with a transfer angle of $\theta$ for a qubit chain with length $N$ are \cite{S_Genest2016, S_Vinet2012_Krawtchouk}

\begin{equation} \label{eq:Jn}
J_n =  \begin{cases}\frac{\pi}{2 \tau} \sqrt{
\frac{n (N - n)\left((N-2n)^2-\left(\frac{ \theta}{\pi}\right)^2 \right)}{(N - 1 - 2n)(N +1 - 2n)}} \text{ for N even}
\\ \frac{\pi}{2 \tau} \sqrt{\frac{n (N - n)\left((N-2n)^2-\left(\frac{\theta}{\pi}- 1\right)^2 \right)}{(N-2n)^2}} \text{ for $N$ odd},
\end{cases}
\end{equation}

\begin{equation} \label{eq:Dn}
\Delta_n = 
\begin{cases}
0  \text{ for $N$ even}
\\
\frac{\pi}{ 2\tau} \frac{\left(\frac{ \theta}{\pi}-1\right) N}{2}  \left( \frac{1}{2n - N}  - \frac{1}{2n - 2 - N} \right) \text{ for $N$ odd}.
\end{cases}
\end{equation}

As the coupling strengths and detunings can not be chosen arbitrarily large in experimental realizations, they determine the speed of our operation. 
For perfect state transfer, when $\theta = \pi$, the formulas simplify to the known result of 
\begin{equation} \label{eq:PST}
J_n = \frac{\pi}{2 \tau} \sqrt{n (N - n)},
\end{equation}
with constant $\Delta_n$. Hence, the biggest required coupling is in the middle of the chain, where $J_{N/2}= \frac{\pi}{2 \tau} \frac{N}{2}$ ($J_{(N+1)/2}=  \frac{\pi}{4 \tau} \sqrt{N^2 -1}$) for $N$ even (odd).
Then, for a maximum coupling $J_\text{max}$ perfect state transfer would be implemented in $\tau=\frac{N\pi}{4 J_\text{max}}$ ($\tau=\frac{\sqrt{N^2 -1}\pi}{4 J_\text{max}}$) for $N$ even (odd).

For FST we analyze the behaviour as we reduce the transfer angle $\theta$ from $\theta=\pi$ in the perfect state transfer case: for $N$ odd we have $\frac{\text d J_n}{\text d \theta} \geq 0$ indicating a speed-up for smaller angles; for $N$ even $\frac{\text d J_n}{\text d \theta} \leq 0 \, \forall n \neq \frac{N}{2}$ and $\frac{\text d J_n}{\text d \theta} \geq 0$ for $ n = \frac{N}{2}$ resulting in increased gate times at small angles. In fact the operation time is the longest for $\theta \rightarrow 0$ (see Fig.~\ref{fig:JD}). Upper bounds for the minimum gate times $\tau$ holding for all $\theta$ are given, in terms of the highest coupling $J_\text{max}$, as
\begin{equation} \label{Jn odd}
\tau \leq 
\begin{cases}
\frac{\pi}{2 \sqrt{3} J_\text{max}} \sqrt{N^2-4}\text{ for } N \text{even} \\
\frac{\pi}{4 J_\text{max}} \sqrt{N^2 -1} \text{ for } N \text{ odd}.
\end{cases}
\end{equation}
For $N$ odd the required range of detunings is:
\begin{equation}
    \Delta_{\text{max}} -  \Delta_{\text{min}}= \frac{N (\pi -\theta)}{3 \tau}.
\end{equation}

\section{S-II. Details on the mapping between \texorpdfstring{$H_N$}{Lg} and \texorpdfstring{$G_N$}{Lg}}\label{sup:mappGH}

The single-excitation manifold matrix elements of $U_N = \exp(-i \tau H_N)$ and $K_N = \exp(-i \frac{\theta}{2} G_N)$ are
\begin{equation}
\begin{split}
    \bra{n}U_N\ket{m} &= e^{-i\phi} \left(\cos(\frac{\theta}{2}) \delta_{n,m} - i \sin(\frac{\theta}{2}) \delta_{n, N+1-m} \right)
\\
    \bra{n}K_N\ket{m} &= \cos(\frac{\theta}{2}) \delta_{n,m} - i \sin(\frac{\theta}{2}) \delta_{n, N+1-m},
\end{split}
\end{equation}
with the special case of the middle qubit when $N$ odd
\begin{equation}
\begin{split}
    \bra{\frac{N+1}{2}}U_N\ket{\frac{N+1}{2}} &= e^{-i\left(\phi +\theta/2\right)}
    \\
    \bra{\frac{N+1}{2}}K_N\ket{\frac{N+1}{2}} &= 1.
\end{split}
\end{equation}
These matrix elements can be aligned by the local unitary rotation $U_\text z = \exp(i \phi H_\text z)$, with
\begin{equation}
H_{\text z} = \begin{cases}
\sum_n  \sigma_n^+ \sigma_n^-   \text{ for $N$ even}
\\ \sum_n  \sigma_n^+ \sigma_n^-  + \frac{\theta}{2\phi} \sigma_{\frac{N+1}{2}}^+ \sigma_{\frac{N+1}{2}}^-  \text{ for $N$ odd,}
\end{cases}
\end{equation}
such that the equivalence $K_N = U_N U_\text z$ is shown to hold in the single-excitation manifold.
To show this equivalence in all excitation manifolds, we use the 
Jordan-Wigner transformation with fermionic annihilation operators $a_n = -(\otimes_{k=1}^{n-1} Z_k) \otimes \sigma^-_n$.
The Jordan-Wigner transformation of the Hamiltonians $H_N$, $G_N$ and $H_\text z$ is given by
\begin{equation}\label{HNFermionic}
 H_N^{\text{F}} = 
 \sum_{n=1}^{N-1} \left( J_n a_n^\dagger a_{n+1} + \text{h.c.} \right) + \sum_{n=1}^N  \Delta_n a^\dagger_n a_n,
\end{equation}
\begin{equation} \label{GN_fermi}
\begin{split}
 G_N^{\text{F}} &=  a_1^\dagger \otimes I_2 \otimes I_3 \otimes \dots{} \otimes I_{N-2}\otimes I_{N-1} \otimes a_N   
\\
&+ \, I_1\otimes a_2^\dagger\otimes I_3 \otimes \dots{}\otimes I_{N-2} \otimes a_{N-1} \otimes I_N
\\
&+ \dots{} + \mathrm{h.c.},
\end{split}
\end{equation}
and
\begin{equation}
H_{\text z}^{\text{F}} = \begin{cases}
\sum_n  a_n^\dagger a_n   \text{ for $N$ even}
\\ \sum_n a_n^\dagger a_n  + \frac{\theta}{2\phi} a_{\frac{N+1}{2}}^\dagger a_{\frac{N+1}{2}}  \text{ for $N$ odd.}
\end{cases}
\end{equation}
Since these operators are all quadratic in the fermionic creation and annihilation operators, they describe non-interacting fermions \cite{S_Terhal2002, S_Kay2010} and their dynamics are fully determined in the single excitation manifold \cite{S_Cappellaro2011}. Therefore, $K_N = U_N U_\text z$, i.e.\ Eq.~\eqref{eq:mapping} of the main text, holds in all excitation manifolds.
\section{S-III. Decomposition of the FST gate into two-qubit gates}\label{sup:Decomp}

\begin{figure}[t]
    \includegraphics[width=0.98\linewidth]{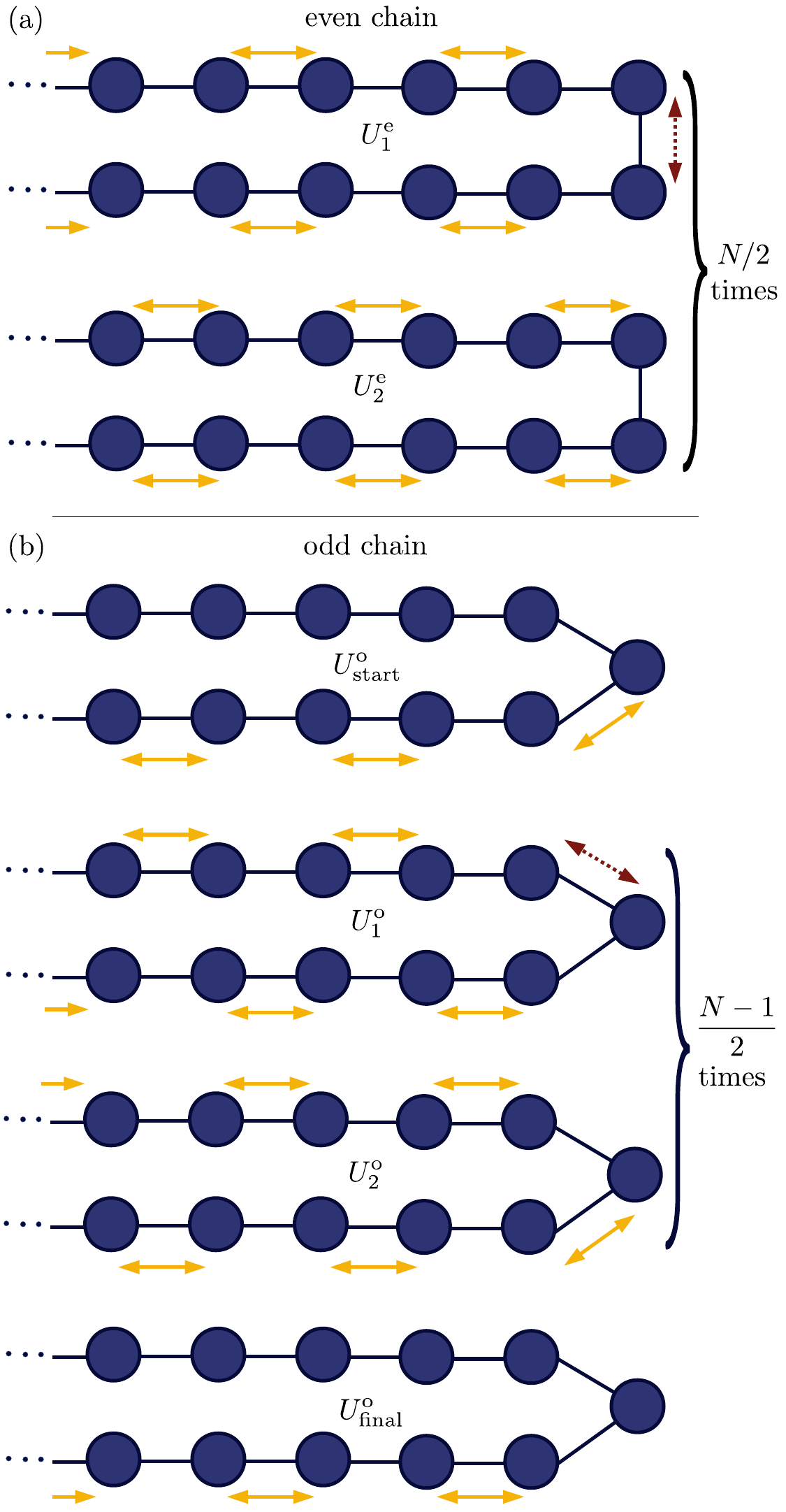}
    \caption{Building blocks of the decomposition of the FST gate into two qubit gates. The chain is folded in half at the middle forming a ladder. Yellow arrows symbolize FSWAP gates. Red dotted arrows symbolize an $\text{iSWAP}(-\theta)$ gate. (a) For $N$ even the FST gate is decomposed by $N/2$ applications of $U^\text{e}_1 U^\text{e}_2$.
    (b) For $N$ odd the FST gate is decomposed by a single application of $U^\text{o}_{\text{start}}$ followed by $(N-1)/2$ applications of $U^\text{o}_1 U^\text{o}_2$ and a final application of $U^\text{o}_{\text{final}}$.
    }\label{fig:decomp}
\end{figure}

The decomposition of the FST gate uses the two qubit gates
\begin{equation}
\begin{aligned}
\text{iSWAP}(\theta)&=\left(
\begin{array}{cccc}
 1 & 0 & 0 & 0 \\
 0 & \cos (\theta ) & i \sin (\theta ) & 0 \\
 0 & i \sin (\theta ) & \cos (\theta ) & 0 \\
 0 & 0 & 0 & 1 \\
\end{array}
\right),\\
\text{FSWAP}&=\left(
\begin{array}{cccc}
 1 & 0 & 0 & 0 \\
 0 & 0 & 1 & 0 \\
 0 & 1 & 0 & 0 \\
 0 & 0 & 0 & -1 \\
\end{array}
\right).\label{eq:Swaps}
\end{aligned}
\end{equation}
Here FSWAP is the fermionic swap gate, which is widely used in the simulation of fermionic system \cite{S_Cade2020}
as it captures the minus sign collected during the exchange of two fermions. It can be decomposed by an iSWAP gate with additional single-qubit $Z$ gates according to
$\text{FSWAP}=Z^{\frac{\pi}{2}}_1 Z^{\frac{\pi}{2}}_2\text{iSWAP}(-\pi)$.
For $N$ even, the FST gate is decomposed similar as in \cite{S_Cade2020} by $N/2$ applications of the unitaries $U^\text{e}_1 U^\text{e}_2$ as defined in Fig.~\ref{fig:decomp}(a). Since for $N > 4$ both $U^\text{e}_1$ and $U^\text{e}_2$ include full FSWAP gates, the decomposition takes time $N \pi/(2 J_{\text{max}})$ independent of $\theta$. Comparing with Sec.~\hyperref[sup:Hparas]{S-I} we see that the direct implementation is at least a factor of $\sqrt{3}$ faster than its decomposition. For $\theta = \pi$ the gate takes half the decomposition's time. The total amount of gates required is $(N^2/2 - N)$ FSWAP gates and $N/2$ iSWAP gates. Since the gate count grows quadratically in $N$, coherent errors are also expected to grow with $\exp(N^2)$. 

For $N$ odd the gate can be decomposed in a single application of $U^\text{o}_{\text{start}}$ followed by $(N-1)/2$ applications of $U^\text{o}_1 U^\text{o}_2$ and a final application of $U^\text{o}_{\text{final}}$ as defined in Fig.~\ref{fig:decomp}(b). For $N>3$ the whole decomposition takes time $(N+1) \pi/(2 J_{\text{max}})$. This is at least twice as long as the direct implementation assuming the gate speed is limited by $J_{\text{max}}$ and not the available detuning range. The total amount of gates needed is $(N-1)^2/2$ FSWAPs and $(N-1)/2$ iSWAPs and also grows quadratic with $N$.

While we don't prove the optimality of these decompositions there has been considerable effort to find efficient decompositions in the case of $N$ even to enable Trotter simulation of a 2D-Fermi-Hubbard model \cite{S_Cade2020}. To the authors' knowledge, no faster decomposition has been found so far.

\section{S-IV. Superconducting three-qubit chain simulation }\label{sup:Eff_H_der}
We simulate the three-qubit chain in a superconducting architecture using the q-optimize software package \cite{S_Wittler2021}. Three fixed-frequency transmons \cite{S_Koch2007} are coupled with two flux tunable coupler transmons \cite{S_Mckay2016} in the dispersive regime (see Fig.~\ref{fig:sim_res}(a) of the main text). The qubits are modeled as Duffing oscillators with three energy levels each. The system Hamiltonian is given by:
\begin{equation}\label{eq:Hduff}
\begin{aligned}
    H &= \sum_{i \in 1,2,3} \left(\omega_i b^\dagger_i b_i + \frac{\alpha_i}{2} b^\dagger_i b^\dagger_i b_i b_i\right) 
    \\& + \sum_{i \in c1, c2} \left(\omega_{i}(\Phi_{i}) b^\dagger_i b_i + \frac{\alpha_i}{2} b^\dagger_i b^\dagger_i b_i b_i\right) 
    \\& - \sum_{i \in 1,2} g_{i,c1} (b^\dagger_i - b_i) (b^\dagger_{c1} - b_{c1})
    \\& - \sum_{i \in 2,3} g_{i,c2} (b^\dagger_i - b_i) (b^\dagger_{c2} - b_{c2})
    \\& -g_{12} (b^\dagger_1 - b_1) (b^\dagger_{2} - b_{2})
    \\& -g_{23} (b^\dagger_2 - b_2) (b^\dagger_{3} - b_{3}),
\end{aligned}
\end{equation}
with 1, 2, 3 ($c1$, $c2$) being the qubit (coupler) indices, $b_i$ ($b^\dagger_i$) bosonic annihilation (creation) operators, $\omega_i$ ($\omega_{ci}$) the bare frequencies of the qubits (couplers),  $\alpha_i$ ($\alpha_{ci}$) the anharmonicities of the qubits (couplers), $g_{i, cj}$ the coupling between qubit $i$ and coupler $j$, and  $ |g_{i, j}|  \ll |g_{i, cj}|$ the direct coupling between two neighboring qubits.
The frequency of coupler $ci$ is modulated with the external flux $\Phi_{ci}$.
The chosen Hamiltonian parameters are summarized in Table~\ref{paras_table}.  

By periodically modulating the coupler frequencies, effective couplings between neighboring qubits arise. High fidelity iSWAP gates have been implemented this way in experiment \cite{S_Mckay2016}.  While adjusting the flux drive amplitudes changes the strength of the effective coupling $J$, adjusting the frequency of the drives introduces the needed detuning $\Delta$ (see Sec.~\hyperref[sec:SV]{S-V}). Therefore, for proper choice of drive frequencies and amplitudes the effective Hamiltonian $H_3$ in Eq.~\eqref{eq:H_3} with conditions Eqs.~\eqref{eq:Jn} and \eqref{eq:Dn} can be realized. 
Note that the chosen architecture can easily be extended to longer qubit chains.
As the interactions are mediated by parametric drives, the gate is compatible with single-qubit virtual $Z$ gates \cite{S_Mckay2017}, which are assumed to be perfect in the following.

Because of their low anharmonicity, coupled transmons suffer from $ZZ$-interactions that result in a frequency shift of states with multiple excitations. As our gate relies on states in the single and second-excitation manifold having the same energy level spacing, these are detrimental to the fidelity of the gate. However, for our choice of couplings there exist three distinct coupler frequencies each, that result in the $ZZ$-interaction for neighboring qubits being zero. We bias the couplers at the second zero point, where we have sufficient effective coupling between the qubits but are still in the dispersive regime. The couplers asymmetries $d_{ci}$ and their fluxbias-points $\Phi_{ci}^\text{DC}$ are chosen such that the average frequencies of the couplers don't shift substantially during the gate to still operate at the $ZZ$ zero point.

The drive pulse is amplitude modulated according to a flattop Gaussian given by 
\begin{equation}
\begin{aligned}
A_{ci}(t) = \Phi^{\mathrm{A}}_{ci} &\times [1 + \text{erf}(t/\tau_{\text r} - 2)] \\
&\times[1 + \text{erf}((\tau_{\text{final}}-t)/\tau_{\text r} - 2)]/4,
\end{aligned}
\end{equation}
where $\text{erf}$ is the Gauss error function, $\tau_{\text{final}}$ the gate length,  $\tau_{\text r}$ the rise time of the pulse and $\Phi^{\mathrm{A}}_{ci} $ the amplitude of the drive on coupler $ci$. This function is sampled with a finite resolution of $2.4\,\text{GHz}$ to model a realistic arbitrary waveform generator. The resulting envelope is mixed with a local oscillator signal with frequency $\omega_{di}$ such that the total flux experienced by the couplers is 
\begin{equation}
\begin{aligned}
    \Phi_{ci}(t) = \Phi_{ci}^\text{DC} + A_{ci}(t)\cos(\omega_{di}t).
\end{aligned}
\end{equation}
The frequency $\omega_{ci}$ of coupler $ci$ is then modulated according to
\begin{equation} \label{eq:wc}
\begin{aligned}
    \omega_{ci}&(\Phi_{ci}) = \alpha_{ci}+\left(\omega_{ci} - \alpha_{ci}\right)\varphi(\Phi_{ci}),
\end{aligned}
\end{equation}
where
\begin{equation}
\begin{aligned}
    \varphi(\Phi_{ci}) = \sqrt[4]{\cos^2\left(\frac{\pi \Phi_{ci}}{\Phi_0}\right) + d_{ci}^2  \sin^2\left(\frac{\pi \Phi_{ci}}{\Phi_0}\right)},
\end{aligned}
\end{equation}
$\Phi_{ci}$ is the applied flux, $\Phi_0$ is the flux quantum and $0 \leq d_{ci} \leq 1$ describes the asymmetry of the coupler. The rise-time of the flattop Gaussian is fixed and only pulse amplitude and frequency of both pulses are optimized resulting in a total of four optimized parameters. The optimization uses the  L-BFGS-B  algorithm \cite{S_Zhu1997} with gradients calculated by numerical differentiation. For $\theta = \pi$ we use a gate time of $\tau_{\text{final}} = 212\,\text{ns}$ and $\tau_{\text r} = 2\,\text{ns}$. For smaller angles the pulse envelope is scaled according to Eq.~\eqref{eq:t3}, i.e.\ by solving the $N=3$ case of Eq.~\eqref{eq:Jn} for~$\tau$. 

\begin{figure}[t]
    \includegraphics[width=0.98\linewidth]{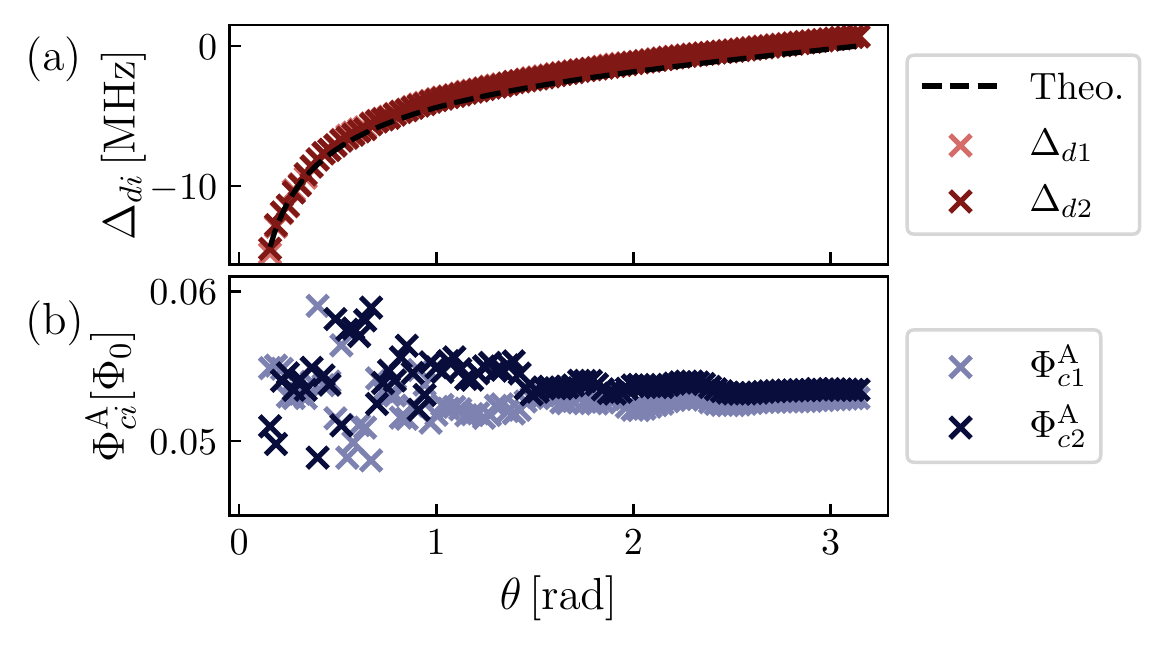}
    \caption{Optimized parameters for the achieved fidelities shown in Fig.~\ref{fig:sim_res} of the main text. (a) Detuning of both drives (markers lie exactly over each other) and theoretical prediction (dashed line). The detuning is slightly shifted from the prediction because of AC-stark shifts.
    (b) Flux-drive amplitude of the flux at the first (bright blue) and second (dark blue) coupler in units of the flux quantum $\Phi_0$. No significant increase in drive amplitude is needed for small angles.}\label{fig:sim_paras} 
\end{figure}

The infidelities of the optimized pulses are shown in Fig.~\ref{fig:sim_res}(b) of the main text and are below $10^{-3}$ for all $\theta$ indicating that the gate will be coherence limited.
The drive detuning $\Delta$ closely follows the theoretical prediction with small deviations likely caused by AC-stark shifts induced by the drive [see Fig.~\ref{fig:sim_paras}(a)]. The expected speed-up for smaller angles predicted in Eq.~\eqref{eq:t3} is achieved without a significant increase in drive amplitude [see Fig.~\ref{fig:sim_paras}(b)]. At small angles, the drive amplitudes change slightly to minimize leakage.

\begin{table}[t]
\begin{tabular}{ c c } 
 \hline
 \hline
Parameter & Value  \\ 
 \hline
$\omega_1/2\pi$ & $5.05\,$GHz  \\ 
$\omega_2/2\pi$ & $5.00\,$GHz  \\ 
$\omega_3/2\pi$ & $5.075\,$GHz  \\
$\omega_{c1}/2\pi$ (at $\Phi_{c1}$) & $6.086\,$GHz  \\
$\omega_{c2}/2\pi$ (at $\Phi_{c2}$) & $6.106\,$GHz  \\
$\alpha_1/2\pi$, $\alpha_2/2\pi$, $\alpha_3/2\pi$ & $-300\,$MHz\\ 
$\alpha_{c1}/2\pi$, $\alpha_{c2}/2\pi$ & $-350\,$MHz\\
$\Phi_{c1}^\text{DC}$, $\Phi_{c2}^\text{DC}$ & $0.3  \Phi_0$\\ 
$d_{c1}$, $d_{c2}$ & $ 0.5$\\ 
$g_{1,c1}/2\pi, g_{2,c2}/2\pi$ & $100\,$MHz  \\ 
$g_{2,c1}/2\pi, g_{3,c2}/2\pi$ & $-100\,$MHz  \\ 
$g_{1,2}/2\pi, g_{2,3}/2\pi$ & $-6.6\,$MHz  \\ 
 \hline 
 \hline
\end{tabular}\caption{\label{paras_table} Hamiltonian parameters used in simulation.}
\end{table}

\section{S-V. Effective Hamiltonian in driven three-qubit chain}\label{sec:SV}

We start from the full Hamiltonian of the chain given in Eq.~\eqref{eq:Hduff}.
If the couplers are in the dispersive regime $\left(\left|\frac{g_{i,cj}} {\omega_{ci} - \omega_{j}} \right|\ll 1\right)$, they decouple from the dynamics and the Hamiltonian can be simplified similar as in \cite{S_Sete2021} by Schrieffer-Wolff transformation. Since we consider transmons we also assume $|\alpha_j| \ll |{\omega_{ci} - \omega_{j}}|$. We obtain
\begin{equation}
\begin{aligned}
    H_{\text{SWT}} &= \sum_{i \in 1,2,3} \left(\tilde \omega_i b^\dagger_i b_i + \frac{\alpha_i}{2} b^\dagger_i b^\dagger_i b_i b_i\right) 
    \\&+ \tilde g_{1,2} (b^\dagger_1  b_2 + \text{h.c.}) + \tilde g_{2,3} (b^\dagger_2  b_3 + \text{h.c.}),
\end{aligned}
\end{equation}
where 
\begin{equation}
\begin{aligned}
\tilde \omega_1 &= \omega_1 - \frac{g_{1, c1}^2}{\Delta_{1, c1}} - \frac{g_{1, c1}^2}{\Sigma_{1, c1}},\\
\tilde \omega_2 &= \omega_2 - \frac{g_{2, c1}^2}{\Delta_{2, c1}} - \frac{g_{2, c1}^2}{\Sigma_{2, c1}} - \frac{g_{2, c2}^2}{\Delta_{2, c2}} - \frac{g_{2, c2}^2}{\Sigma_{2, c2}}, \\
\tilde \omega_3 &= \omega_3 - \frac{g_{3 c2}^2}{\Delta_{3, c2}} - \frac{g_{3, c2}^2}{\Sigma_{3, c2}},
\end{aligned}
\end{equation}
and
\begin{equation}
\begin{aligned}
\tilde g_{i,j}(\Phi) = g_{i,j} &- \frac{g_{i,ci} g_{j, ci}}{2}\sum_{n \in i, j} \left(\frac{1}{\Delta_{n,ci}(\Phi)} + \frac{1}{\Sigma_{n, ci}(\Phi)} \right), \\
\Delta_{n,ci}(\Phi) &= \omega_{ci}(\Phi) - \omega_n, \\
\Sigma_{n,ci}(\Phi) &= \omega_{ci}(\Phi) + \omega_n.
\end{aligned}
\end{equation}
If we periodically drive the flux through the coupler $i$ with amplitude $A_{di}$ and frequency $\omega_{di}$, we can expand $\tilde g_{i,j}\left(\Phi(t)\right)$ in a Fourier series with coefficients $\bar g_{i,j}^{(n)}$: 
\begin{equation}
\tilde g_{i,j}(A_{di} \cos(\omega_{di} t)) = \sum_{n=-\infty}^\infty \bar g_{i,j}^{(n)} \exp(i n \omega_{di} t).
\end{equation}

Assuming $\omega_{d1} \approx |\tilde \omega_1 -  \tilde \omega_2|$ and $\omega_{d2} \approx |\tilde \omega_2 -  \tilde \omega_3|$ we can neglect higher sidebands and only consider $\bar g_{i,j}^{(1)}$. Restricting the Hamiltonian to two level systems and going to a rotating frame with a unitary transformation given by 
\begin{equation}
\begin{aligned}
U = \exp\Big(i&\,t\,\big[(\tilde \omega_2 + \omega_{d1}) \sigma^+_1 \sigma^-_1 + \tilde \omega_2 \sigma^+_2 \sigma^-_2
\\&+ (\tilde \omega_2 + \omega_{d2}) \sigma^+_3 \sigma^-_3 \big]\Big),
\end{aligned}
\end{equation}
we end up with the effective Hamiltonian
\begin{equation}
\begin{aligned}
    H_{\text{RF}} &= \Delta_{d1}\sigma^+_1 \sigma^-_1 + \Delta_{d2} \sigma^+_3 \sigma^-_3
    \\&+ \bar g_{1,2}^{(1)} (\sigma^+_1  \sigma^-_2 + \text{h.c.}) 
    \\&+ \bar g_{2,3}^{(1)} (\sigma^+_2  \sigma^-_3 + \text{h.c.}),
\end{aligned}
\end{equation}
where 
\begin{equation}
\begin{aligned}
\Delta_{d1} &= (\tilde \omega_{1} -  \tilde \omega_{2}) - \omega_{d1},\\
\Delta_{d2} &= (\tilde \omega_{3} -  \tilde \omega_{2}) - \omega_{d2}.
\end{aligned}
\end{equation}
By choosing $\Delta_{d1} = \Delta_{d2} = \Delta$ by adjusting the drive frequencies,  setting $\bar g_{1,2}^{(1)} = \bar g_{2,3}^{(1)} = J$ by adjusting the flux drive amplitudes and going into the frame rotating at frequency $\Delta$ we get the effective Hamiltonian
\begin{equation}\label{eq:H_3}
    H_3 = -\Delta \sigma_2^+ \sigma_2^- + J \left( \sigma_1^+ \sigma_2^- + \sigma_2^+ \sigma_3^- + \text{h.c.} \right),
\end{equation}
which is exactly the Hamiltonian needed for FST.
The unitary 
\begin{equation}
   K_3 = \exp(-i \theta \left(\sigma_1^+ Z_2 \sigma_3^- + \text{h.c.} \right))
\end{equation}
is realized by setting 
\begin{equation}
\Delta = \frac{2 J (\pi - \theta )}{\sqrt{(\pi -\frac{\theta}{2} ) \theta }},
\end{equation}
evolving under $H_3$ for 
\begin{equation}\label{eq:t3}
    \tau = \frac{\sqrt{(\pi -\frac{\theta}{2} ) \theta }}{J}
\end{equation} and applying single-qubit $Z$-rotations.
These parameters where also found in \cite{S_Burkhart2021}, where the middle qubit is replaced by a transmission line  initially in the ground state and the other two qubits by resonators.
The single-qubit $Z$-rotations needed are $U_{\text Z_{1/3}} = \exp\left(i \frac{\theta}{2} \sigma_{1/3}^+ \sigma_{1/3}^-  \right)$ and $U_{\text Z_2} = \exp\left(i  \theta \sigma_2^+ \sigma_2^- \right)$.

\input{supp.bbl}
\end{document}

%% file: main.bbl

%% file: main.bbl
\begin{thebibliography}{68}%
\makeatletter
\providecommand \@ifxundefined [1]{%
 \@ifx{#1\undefined}
}%
\providecommand \@ifnum [1]{%
 \ifnum #1\expandafter \@firstoftwo
 \else \expandafter \@secondoftwo
 \fi
}%
\providecommand \@ifx [1]{%
 \ifx #1\expandafter \@firstoftwo
 \else \expandafter \@secondoftwo
 \fi
}%
\providecommand \natexlab [1]{#1}%
\providecommand \enquote  [1]{``#1''}%
\providecommand \bibnamefont  [1]{#1}%
\providecommand \bibfnamefont [1]{#1}%
\providecommand \citenamefont [1]{#1}%
\providecommand \href@noop [0]{\@secondoftwo}%
\providecommand \href [0]{\begingroup \@sanitize@url \@href}%
\providecommand \@href[1]{\@@startlink{#1}\@@href}%
\providecommand \@@href[1]{\endgroup#1\@@endlink}%
\providecommand \@sanitize@url [0]{\catcode `\\12\catcode `\$12\catcode
  `\&12\catcode `\#12\catcode `\^12\catcode `\_12\catcode `\%12\relax}%
\providecommand \@@startlink[1]{}%
\providecommand \@@endlink[0]{}%
\providecommand \url  [0]{\begingroup\@sanitize@url \@url }%
\providecommand \@url [1]{\endgroup\@href {#1}{\urlprefix }}%
\providecommand \urlprefix  [0]{URL }%
\providecommand \Eprint [0]{\href }%
\providecommand \doibase [0]{https://doi.org/}%
\providecommand \selectlanguage [0]{\@gobble}%
\providecommand \bibinfo  [0]{\@secondoftwo}%
\providecommand \bibfield  [0]{\@secondoftwo}%
\providecommand \translation [1]{[#1]}%
\providecommand \BibitemOpen [0]{}%
\providecommand \bibitemStop [0]{}%
\providecommand \bibitemNoStop [0]{.\EOS\space}%
\providecommand \EOS [0]{\spacefactor3000\relax}%
\providecommand \BibitemShut  [1]{\csname bibitem#1\endcsname}%
\let\auto@bib@innerbib\@empty
%</preamble>
\bibitem [{\citenamefont {Kim}\ \emph {et~al.}(2021)\citenamefont {Kim},
  \citenamefont {Wood}, \citenamefont {Yoder}, \citenamefont {Merkel},
  \citenamefont {Gambetta}, \citenamefont {Temme},\ and\ \citenamefont
  {Kandala}}]{Kim2021}%
  \BibitemOpen
  \bibfield  {author} {\bibinfo {author} {\bibfnamefont {Y.}~\bibnamefont
  {Kim}}, \bibinfo {author} {\bibfnamefont {C.~J.}\ \bibnamefont {Wood}},
  \bibinfo {author} {\bibfnamefont {T.~J.}\ \bibnamefont {Yoder}}, \bibinfo
  {author} {\bibfnamefont {S.~T.}\ \bibnamefont {Merkel}}, \bibinfo {author}
  {\bibfnamefont {J.~M.}\ \bibnamefont {Gambetta}}, \bibinfo {author}
  {\bibfnamefont {K.}~\bibnamefont {Temme}},\ and\ \bibinfo {author}
  {\bibfnamefont {A.}~\bibnamefont {Kandala}},\ }\bibfield  {title} {\bibinfo
  {title} {Scalable error mitigation for noisy quantum circuits produces
  competitive expectation values},\ }\href {https://arxiv.org/abs/2108.09197}
  {\bibfield  {journal} {\bibinfo  {journal} {arXiv:2108.09197}\ } (\bibinfo
  {year} {2021})}\BibitemShut {NoStop}%
\bibitem [{\citenamefont {Arute}\ \emph {et~al.}(2020)\citenamefont {Arute},
  \citenamefont {Arya}, \citenamefont {Babbush}, \citenamefont {Bacon},
  \citenamefont {Bardin}, \citenamefont {Barends}, \citenamefont {Boixo},
  \citenamefont {Broughton}, \citenamefont {Buckley}, \citenamefont {Buell}
  \emph {et~al.}}]{Arute2020}%
  \BibitemOpen
  \bibfield  {author} {\bibinfo {author} {\bibfnamefont {F.}~\bibnamefont
  {Arute}}, \bibinfo {author} {\bibfnamefont {K.}~\bibnamefont {Arya}},
  \bibinfo {author} {\bibfnamefont {R.}~\bibnamefont {Babbush}}, \bibinfo
  {author} {\bibfnamefont {D.}~\bibnamefont {Bacon}}, \bibinfo {author}
  {\bibfnamefont {J.~C.}\ \bibnamefont {Bardin}}, \bibinfo {author}
  {\bibfnamefont {R.}~\bibnamefont {Barends}}, \bibinfo {author} {\bibfnamefont
  {S.}~\bibnamefont {Boixo}}, \bibinfo {author} {\bibfnamefont
  {M.}~\bibnamefont {Broughton}}, \bibinfo {author} {\bibfnamefont {B.~B.}\
  \bibnamefont {Buckley}}, \bibinfo {author} {\bibfnamefont {D.~A.}\
  \bibnamefont {Buell}}, \emph {et~al.},\ }\bibfield  {title} {\bibinfo {title}
  {Hartree-fock on a superconducting qubit quantum computer},\ }\href
  {https://doi.org/10.1126/science.abb9811} {\bibfield  {journal} {\bibinfo
  {journal} {Science}\ }\textbf {\bibinfo {volume} {369}},\ \bibinfo {pages}
  {1084} (\bibinfo {year} {2020})}\BibitemShut {NoStop}%
\bibitem [{\citenamefont {Nam}\ \emph {et~al.}(2020)\citenamefont {Nam},
  \citenamefont {Chen}, \citenamefont {Pisenti}, \citenamefont {Wright},
  \citenamefont {Delaney}, \citenamefont {Maslov}, \citenamefont {Brown},
  \citenamefont {Allen}, \citenamefont {Amini}, \citenamefont {Apisdorf} \emph
  {et~al.}}]{Nam2020}%
  \BibitemOpen
  \bibfield  {author} {\bibinfo {author} {\bibfnamefont {Y.}~\bibnamefont
  {Nam}}, \bibinfo {author} {\bibfnamefont {J.-S.}\ \bibnamefont {Chen}},
  \bibinfo {author} {\bibfnamefont {N.~C.}\ \bibnamefont {Pisenti}}, \bibinfo
  {author} {\bibfnamefont {K.}~\bibnamefont {Wright}}, \bibinfo {author}
  {\bibfnamefont {C.}~\bibnamefont {Delaney}}, \bibinfo {author} {\bibfnamefont
  {D.}~\bibnamefont {Maslov}}, \bibinfo {author} {\bibfnamefont {K.~R.}\
  \bibnamefont {Brown}}, \bibinfo {author} {\bibfnamefont {S.}~\bibnamefont
  {Allen}}, \bibinfo {author} {\bibfnamefont {J.~M.}\ \bibnamefont {Amini}},
  \bibinfo {author} {\bibfnamefont {J.}~\bibnamefont {Apisdorf}}, \emph
  {et~al.},\ }\bibfield  {title} {\bibinfo {title} {Ground-state energy
  estimation of the water molecule on a trapped-ion quantum computer},\ }\href
  {https://doi.org/10.1038/s41534-020-0259-3} {\bibfield  {journal} {\bibinfo
  {journal} {npj Quantum Information}\ }\textbf {\bibinfo {volume} {6}},\
  \bibinfo {pages} {1} (\bibinfo {year} {2020})}\BibitemShut {NoStop}%
\bibitem [{\citenamefont {Hempel}\ \emph {et~al.}(2018)\citenamefont {Hempel},
  \citenamefont {Maier}, \citenamefont {Romero}, \citenamefont {McClean},
  \citenamefont {Monz}, \citenamefont {Shen}, \citenamefont {Jurcevic},
  \citenamefont {Lanyon}, \citenamefont {Love}, \citenamefont {Babbush},
  \citenamefont {Aspuru-Guzik}, \citenamefont {Blatt},\ and\ \citenamefont
  {Roos}}]{Hempel2018}%
  \BibitemOpen
  \bibfield  {author} {\bibinfo {author} {\bibfnamefont {C.}~\bibnamefont
  {Hempel}}, \bibinfo {author} {\bibfnamefont {C.}~\bibnamefont {Maier}},
  \bibinfo {author} {\bibfnamefont {J.}~\bibnamefont {Romero}}, \bibinfo
  {author} {\bibfnamefont {J.}~\bibnamefont {McClean}}, \bibinfo {author}
  {\bibfnamefont {T.}~\bibnamefont {Monz}}, \bibinfo {author} {\bibfnamefont
  {H.}~\bibnamefont {Shen}}, \bibinfo {author} {\bibfnamefont {P.}~\bibnamefont
  {Jurcevic}}, \bibinfo {author} {\bibfnamefont {B.~P.}\ \bibnamefont
  {Lanyon}}, \bibinfo {author} {\bibfnamefont {P.}~\bibnamefont {Love}},
  \bibinfo {author} {\bibfnamefont {R.}~\bibnamefont {Babbush}}, \bibinfo
  {author} {\bibfnamefont {A.}~\bibnamefont {Aspuru-Guzik}}, \bibinfo {author}
  {\bibfnamefont {R.}~\bibnamefont {Blatt}},\ and\ \bibinfo {author}
  {\bibfnamefont {C.~F.}\ \bibnamefont {Roos}},\ }\bibfield  {title} {\bibinfo
  {title} {Quantum chemistry calculations on a trapped-ion quantum simulator},\
  }\href {https://doi.org/10.1103/PhysRevX.8.031022} {\bibfield  {journal}
  {\bibinfo  {journal} {Phys. Rev. X}\ }\textbf {\bibinfo {volume} {8}},\
  \bibinfo {pages} {031022} (\bibinfo {year} {2018})}\BibitemShut {NoStop}%
\bibitem [{\citenamefont {Cirac}\ and\ \citenamefont
  {Zoller}(2012)}]{Cirac2012}%
  \BibitemOpen
  \bibfield  {author} {\bibinfo {author} {\bibfnamefont {J.~I.}\ \bibnamefont
  {Cirac}}\ and\ \bibinfo {author} {\bibfnamefont {P.}~\bibnamefont {Zoller}},\
  }\bibfield  {title} {\bibinfo {title} {Goals and opportunities in quantum
  simulation},\ }\href {https://doi.org/10.1038/nphys2275} {\bibfield
  {journal} {\bibinfo  {journal} {Nature physics}\ }\textbf {\bibinfo {volume}
  {8}},\ \bibinfo {pages} {264} (\bibinfo {year} {2012})}\BibitemShut {NoStop}%
\bibitem [{\citenamefont {Bharti}\ \emph {et~al.}(2022)\citenamefont {Bharti},
  \citenamefont {Cervera-Lierta}, \citenamefont {Kyaw}, \citenamefont {Haug},
  \citenamefont {Alperin-Lea}, \citenamefont {Anand}, \citenamefont {Degroote},
  \citenamefont {Heimonen}, \citenamefont {Kottmann}, \citenamefont {Menke},
  \citenamefont {Mok}, \citenamefont {Sim}, \citenamefont {Kwek},\ and\
  \citenamefont {Aspuru-Guzik}}]{Bharti2021}%
  \BibitemOpen
  \bibfield  {author} {\bibinfo {author} {\bibfnamefont {K.}~\bibnamefont
  {Bharti}}, \bibinfo {author} {\bibfnamefont {A.}~\bibnamefont
  {Cervera-Lierta}}, \bibinfo {author} {\bibfnamefont {T.~H.}\ \bibnamefont
  {Kyaw}}, \bibinfo {author} {\bibfnamefont {T.}~\bibnamefont {Haug}}, \bibinfo
  {author} {\bibfnamefont {S.}~\bibnamefont {Alperin-Lea}}, \bibinfo {author}
  {\bibfnamefont {A.}~\bibnamefont {Anand}}, \bibinfo {author} {\bibfnamefont
  {M.}~\bibnamefont {Degroote}}, \bibinfo {author} {\bibfnamefont
  {H.}~\bibnamefont {Heimonen}}, \bibinfo {author} {\bibfnamefont {J.~S.}\
  \bibnamefont {Kottmann}}, \bibinfo {author} {\bibfnamefont {T.}~\bibnamefont
  {Menke}}, \bibinfo {author} {\bibfnamefont {W.-K.}\ \bibnamefont {Mok}},
  \bibinfo {author} {\bibfnamefont {S.}~\bibnamefont {Sim}}, \bibinfo {author}
  {\bibfnamefont {L.-C.}\ \bibnamefont {Kwek}},\ and\ \bibinfo {author}
  {\bibfnamefont {A.}~\bibnamefont {Aspuru-Guzik}},\ }\bibfield  {title}
  {\bibinfo {title} {Noisy intermediate-scale quantum algorithms},\ }\href
  {https://doi.org/10.1103/RevModPhys.94.015004} {\bibfield  {journal}
  {\bibinfo  {journal} {Rev. Mod. Phys.}\ }\textbf {\bibinfo {volume} {94}},\
  \bibinfo {pages} {015004} (\bibinfo {year} {2022})}\BibitemShut {NoStop}%
\bibitem [{\citenamefont {Moll}\ \emph {et~al.}(2018)\citenamefont {Moll},
  \citenamefont {Barkoutsos}, \citenamefont {Bishop}, \citenamefont {Chow},
  \citenamefont {Cross}, \citenamefont {Egger}, \citenamefont {Filipp},
  \citenamefont {Fuhrer}, \citenamefont {Gambetta}, \citenamefont {Ganzhorn}
  \emph {et~al.}}]{Moll2018}%
  \BibitemOpen
  \bibfield  {author} {\bibinfo {author} {\bibfnamefont {N.}~\bibnamefont
  {Moll}}, \bibinfo {author} {\bibfnamefont {P.}~\bibnamefont {Barkoutsos}},
  \bibinfo {author} {\bibfnamefont {L.~S.}\ \bibnamefont {Bishop}}, \bibinfo
  {author} {\bibfnamefont {J.~M.}\ \bibnamefont {Chow}}, \bibinfo {author}
  {\bibfnamefont {A.}~\bibnamefont {Cross}}, \bibinfo {author} {\bibfnamefont
  {D.~J.}\ \bibnamefont {Egger}}, \bibinfo {author} {\bibfnamefont
  {S.}~\bibnamefont {Filipp}}, \bibinfo {author} {\bibfnamefont
  {A.}~\bibnamefont {Fuhrer}}, \bibinfo {author} {\bibfnamefont {J.~M.}\
  \bibnamefont {Gambetta}}, \bibinfo {author} {\bibfnamefont {M.}~\bibnamefont
  {Ganzhorn}}, \emph {et~al.},\ }\bibfield  {title} {\bibinfo {title} {Quantum
  optimization using variational algorithms on near-term quantum devices},\
  }\href {https://doi.org/10.1088/2058-9565/aab822} {\bibfield  {journal}
  {\bibinfo  {journal} {Quantum Science and Technology}\ }\textbf {\bibinfo
  {volume} {3}},\ \bibinfo {pages} {030503} (\bibinfo {year}
  {2018})}\BibitemShut {NoStop}%
\bibitem [{\citenamefont {McArdle}\ \emph {et~al.}(2020)\citenamefont
  {McArdle}, \citenamefont {Endo}, \citenamefont {Aspuru-Guzik}, \citenamefont
  {Benjamin},\ and\ \citenamefont {Yuan}}]{Mcardle2020}%
  \BibitemOpen
  \bibfield  {author} {\bibinfo {author} {\bibfnamefont {S.}~\bibnamefont
  {McArdle}}, \bibinfo {author} {\bibfnamefont {S.}~\bibnamefont {Endo}},
  \bibinfo {author} {\bibfnamefont {A.}~\bibnamefont {Aspuru-Guzik}}, \bibinfo
  {author} {\bibfnamefont {S.~C.}\ \bibnamefont {Benjamin}},\ and\ \bibinfo
  {author} {\bibfnamefont {X.}~\bibnamefont {Yuan}},\ }\bibfield  {title}
  {\bibinfo {title} {Quantum computational chemistry},\ }\href
  {https://doi.org/10.1103/RevModPhys.92.015003} {\bibfield  {journal}
  {\bibinfo  {journal} {Rev. Mod. Phys.}\ }\textbf {\bibinfo {volume} {92}},\
  \bibinfo {pages} {015003} (\bibinfo {year} {2020})}\BibitemShut {NoStop}%
\bibitem [{\citenamefont {Esslinger}(2010)}]{Esslinger2010}%
  \BibitemOpen
  \bibfield  {author} {\bibinfo {author} {\bibfnamefont {T.}~\bibnamefont
  {Esslinger}},\ }\bibfield  {title} {\bibinfo {title} {Fermi-hubbard physics
  with atoms in an optical lattice},\ }\href
  {https://doi.org/10.1146/annurev-conmatphys-070909-104059} {\bibfield
  {journal} {\bibinfo  {journal} {Annual Review of Condensed Matter Physics}\
  }\textbf {\bibinfo {volume} {1}},\ \bibinfo {pages} {129} (\bibinfo {year}
  {2010})}\BibitemShut {NoStop}%
\bibitem [{\citenamefont {Havl\'{\i}\ifmmode~\check{c}\else \v{c}\fi{}ek}\
  \emph {et~al.}(2017)\citenamefont {Havl\'{\i}\ifmmode~\check{c}\else
  \v{c}\fi{}ek}, \citenamefont {Troyer},\ and\ \citenamefont
  {Whitfield}}]{Havl2017}%
  \BibitemOpen
  \bibfield  {author} {\bibinfo {author} {\bibfnamefont {V.~c.~v.}\
  \bibnamefont {Havl\'{\i}\ifmmode~\check{c}\else \v{c}\fi{}ek}}, \bibinfo
  {author} {\bibfnamefont {M.}~\bibnamefont {Troyer}},\ and\ \bibinfo {author}
  {\bibfnamefont {J.~D.}\ \bibnamefont {Whitfield}},\ }\bibfield  {title}
  {\bibinfo {title} {Operator locality in the quantum simulation of fermionic
  models},\ }\href {https://doi.org/10.1103/PhysRevA.95.032332} {\bibfield
  {journal} {\bibinfo  {journal} {Phys. Rev. A}\ }\textbf {\bibinfo {volume}
  {95}},\ \bibinfo {pages} {032332} (\bibinfo {year} {2017})}\BibitemShut
  {NoStop}%
\bibitem [{\citenamefont {Bravyi}\ and\ \citenamefont
  {Kitaev}(2002)}]{Bravyi2002}%
  \BibitemOpen
  \bibfield  {author} {\bibinfo {author} {\bibfnamefont {S.~B.}\ \bibnamefont
  {Bravyi}}\ and\ \bibinfo {author} {\bibfnamefont {A.~Y.}\ \bibnamefont
  {Kitaev}},\ }\bibfield  {title} {\bibinfo {title} {Fermionic quantum
  computation},\ }\href
  {https://doi.org/https://doi.org/10.1006/aphy.2002.6254} {\bibfield
  {journal} {\bibinfo  {journal} {Annals of Physics}\ }\textbf {\bibinfo
  {volume} {298}},\ \bibinfo {pages} {210} (\bibinfo {year}
  {2002})}\BibitemShut {NoStop}%
\bibitem [{\citenamefont {Nielsen}(2005)}]{Nielsen2005}%
  \BibitemOpen
  \bibfield  {author} {\bibinfo {author} {\bibfnamefont {M.~A.}\ \bibnamefont
  {Nielsen}},\ }\bibfield  {title} {\bibinfo {title} {The fermionic canonical
  commutation relations and the {Jordan-Wigner} transform},\ }\href
  {https://futureofmatter.com/assets/fermions_and_jordan_wigner.pdf} {\bibfield
   {journal} {\bibinfo  {journal} {School of Physical Sciences The University
  of Queensland}\ }\textbf {\bibinfo {volume} {59}} (\bibinfo {year}
  {2005})}\BibitemShut {NoStop}%
\bibitem [{\citenamefont {Bruzewicz}\ \emph {et~al.}(2019)\citenamefont
  {Bruzewicz}, \citenamefont {Chiaverini}, \citenamefont {McConnell},\ and\
  \citenamefont {Sage}}]{Bruzewicz2019}%
  \BibitemOpen
  \bibfield  {author} {\bibinfo {author} {\bibfnamefont {C.~D.}\ \bibnamefont
  {Bruzewicz}}, \bibinfo {author} {\bibfnamefont {J.}~\bibnamefont
  {Chiaverini}}, \bibinfo {author} {\bibfnamefont {R.}~\bibnamefont
  {McConnell}},\ and\ \bibinfo {author} {\bibfnamefont {J.~M.}\ \bibnamefont
  {Sage}},\ }\bibfield  {title} {\bibinfo {title} {Trapped-ion quantum
  computing: Progress and challenges},\ }\href
  {https://doi.org/10.1063/1.5088164} {\bibfield  {journal} {\bibinfo
  {journal} {Applied Physics Reviews}\ }\textbf {\bibinfo {volume} {6}},\
  \bibinfo {pages} {021314} (\bibinfo {year} {2019})}\BibitemShut {NoStop}%
\bibitem [{\citenamefont {Clarke}\ and\ \citenamefont
  {Wilhelm}(2008)}]{Clarke2008}%
  \BibitemOpen
  \bibfield  {author} {\bibinfo {author} {\bibfnamefont {J.}~\bibnamefont
  {Clarke}}\ and\ \bibinfo {author} {\bibfnamefont {F.~K.}\ \bibnamefont
  {Wilhelm}},\ }\bibfield  {title} {\bibinfo {title} {Superconducting quantum
  bits},\ }\href {https://doi.org/10.1038/nature07128} {\bibfield  {journal}
  {\bibinfo  {journal} {Nature}\ }\textbf {\bibinfo {volume} {453}},\ \bibinfo
  {pages} {1031} (\bibinfo {year} {2008})}\BibitemShut {NoStop}%
\bibitem [{\citenamefont {Krantz}\ \emph {et~al.}(2019)\citenamefont {Krantz},
  \citenamefont {Kjaergaard}, \citenamefont {Yan}, \citenamefont {Orlando},
  \citenamefont {Gustavsson},\ and\ \citenamefont {Oliver}}]{Krantz2019}%
  \BibitemOpen
  \bibfield  {author} {\bibinfo {author} {\bibfnamefont {P.}~\bibnamefont
  {Krantz}}, \bibinfo {author} {\bibfnamefont {M.}~\bibnamefont {Kjaergaard}},
  \bibinfo {author} {\bibfnamefont {F.}~\bibnamefont {Yan}}, \bibinfo {author}
  {\bibfnamefont {T.~P.}\ \bibnamefont {Orlando}}, \bibinfo {author}
  {\bibfnamefont {S.}~\bibnamefont {Gustavsson}},\ and\ \bibinfo {author}
  {\bibfnamefont {W.~D.}\ \bibnamefont {Oliver}},\ }\bibfield  {title}
  {\bibinfo {title} {A quantum engineer's guide to superconducting qubits},\
  }\href {https://doi.org/10.1063/1.5089550} {\bibfield  {journal} {\bibinfo
  {journal} {Applied Physics Reviews}\ }\textbf {\bibinfo {volume} {6}},\
  \bibinfo {pages} {021318} (\bibinfo {year} {2019})}\BibitemShut {NoStop}%
\bibitem [{\citenamefont {Kloeffel}\ and\ \citenamefont
  {Loss}(2013)}]{Kloeffel2013}%
  \BibitemOpen
  \bibfield  {author} {\bibinfo {author} {\bibfnamefont {C.}~\bibnamefont
  {Kloeffel}}\ and\ \bibinfo {author} {\bibfnamefont {D.}~\bibnamefont
  {Loss}},\ }\bibfield  {title} {\bibinfo {title} {Prospects for spin-based
  quantum computing in quantum dots},\ }\href
  {https://doi.org/10.1146/annurev-conmatphys-030212-184248} {\bibfield
  {journal} {\bibinfo  {journal} {Annu. Rev. Condens. Matter Phys.}\ }\textbf
  {\bibinfo {volume} {4}},\ \bibinfo {pages} {51} (\bibinfo {year}
  {2013})}\BibitemShut {NoStop}%
\bibitem [{\citenamefont {Henriet}\ \emph {et~al.}(2020)\citenamefont
  {Henriet}, \citenamefont {Beguin}, \citenamefont {Signoles}, \citenamefont
  {Lahaye}, \citenamefont {Browaeys}, \citenamefont {Reymond},\ and\
  \citenamefont {Jurczak}}]{Henriet2020}%
  \BibitemOpen
  \bibfield  {author} {\bibinfo {author} {\bibfnamefont {L.}~\bibnamefont
  {Henriet}}, \bibinfo {author} {\bibfnamefont {L.}~\bibnamefont {Beguin}},
  \bibinfo {author} {\bibfnamefont {A.}~\bibnamefont {Signoles}}, \bibinfo
  {author} {\bibfnamefont {T.}~\bibnamefont {Lahaye}}, \bibinfo {author}
  {\bibfnamefont {A.}~\bibnamefont {Browaeys}}, \bibinfo {author}
  {\bibfnamefont {G.-O.}\ \bibnamefont {Reymond}},\ and\ \bibinfo {author}
  {\bibfnamefont {C.}~\bibnamefont {Jurczak}},\ }\bibfield  {title} {\bibinfo
  {title} {Quantum computing with neutral atoms},\ }\href
  {https://doi.org/10.22331/q-2020-09-21-327} {\bibfield  {journal} {\bibinfo
  {journal} {{Quantum}}\ }\textbf {\bibinfo {volume} {4}},\ \bibinfo {pages}
  {327} (\bibinfo {year} {2020})}\BibitemShut {NoStop}%
\bibitem [{\citenamefont {Kok}\ \emph {et~al.}(2007)\citenamefont {Kok},
  \citenamefont {Munro}, \citenamefont {Nemoto}, \citenamefont {Ralph},
  \citenamefont {Dowling},\ and\ \citenamefont {Milburn}}]{Kok2007}%
  \BibitemOpen
  \bibfield  {author} {\bibinfo {author} {\bibfnamefont {P.}~\bibnamefont
  {Kok}}, \bibinfo {author} {\bibfnamefont {W.~J.}\ \bibnamefont {Munro}},
  \bibinfo {author} {\bibfnamefont {K.}~\bibnamefont {Nemoto}}, \bibinfo
  {author} {\bibfnamefont {T.~C.}\ \bibnamefont {Ralph}}, \bibinfo {author}
  {\bibfnamefont {J.~P.}\ \bibnamefont {Dowling}},\ and\ \bibinfo {author}
  {\bibfnamefont {G.~J.}\ \bibnamefont {Milburn}},\ }\bibfield  {title}
  {\bibinfo {title} {Linear optical quantum computing with photonic qubits},\
  }\href {https://doi.org/10.1103/RevModPhys.79.135} {\bibfield  {journal}
  {\bibinfo  {journal} {Reviews of modern physics}\ }\textbf {\bibinfo {volume}
  {79}},\ \bibinfo {pages} {135} (\bibinfo {year} {2007})}\BibitemShut
  {NoStop}%
\bibitem [{\citenamefont {Linke}\ \emph {et~al.}(2017)\citenamefont {Linke},
  \citenamefont {Maslov}, \citenamefont {Roetteler}, \citenamefont {Debnath},
  \citenamefont {Figgatt}, \citenamefont {Landsman}, \citenamefont {Wright},\
  and\ \citenamefont {Monroe}}]{Linke2017}%
  \BibitemOpen
  \bibfield  {author} {\bibinfo {author} {\bibfnamefont {N.~M.}\ \bibnamefont
  {Linke}}, \bibinfo {author} {\bibfnamefont {D.}~\bibnamefont {Maslov}},
  \bibinfo {author} {\bibfnamefont {M.}~\bibnamefont {Roetteler}}, \bibinfo
  {author} {\bibfnamefont {S.}~\bibnamefont {Debnath}}, \bibinfo {author}
  {\bibfnamefont {C.}~\bibnamefont {Figgatt}}, \bibinfo {author} {\bibfnamefont
  {K.~A.}\ \bibnamefont {Landsman}}, \bibinfo {author} {\bibfnamefont
  {K.}~\bibnamefont {Wright}},\ and\ \bibinfo {author} {\bibfnamefont
  {C.}~\bibnamefont {Monroe}},\ }\bibfield  {title} {\bibinfo {title}
  {Experimental comparison of two quantum computing architectures},\ }\href
  {https://doi.org/10.1073/pnas.1618020114} {\bibfield  {journal} {\bibinfo
  {journal} {Proceedings of the National Academy of Sciences}\ }\textbf
  {\bibinfo {volume} {114}},\ \bibinfo {pages} {3305} (\bibinfo {year}
  {2017})}\BibitemShut {NoStop}%
\bibitem [{\citenamefont {Cowtan}\ \emph {et~al.}(2019)\citenamefont {Cowtan},
  \citenamefont {Dilkes}, \citenamefont {Duncan}, \citenamefont {Krajenbrink},
  \citenamefont {Simmons},\ and\ \citenamefont {Sivarajah}}]{Cowtan2019}%
  \BibitemOpen
  \bibfield  {author} {\bibinfo {author} {\bibfnamefont {A.}~\bibnamefont
  {Cowtan}}, \bibinfo {author} {\bibfnamefont {S.}~\bibnamefont {Dilkes}},
  \bibinfo {author} {\bibfnamefont {R.}~\bibnamefont {Duncan}}, \bibinfo
  {author} {\bibfnamefont {A.}~\bibnamefont {Krajenbrink}}, \bibinfo {author}
  {\bibfnamefont {W.}~\bibnamefont {Simmons}},\ and\ \bibinfo {author}
  {\bibfnamefont {S.}~\bibnamefont {Sivarajah}},\ }\bibfield  {title} {\bibinfo
  {title} {On the qubit routing problem},\ }in\ \href
  {https://doi.org/10.4230/LIPIcs.TQC.2019.5} {\emph {\bibinfo {booktitle}
  {14th Conference on the Theory of Quantum Computation, Communication and
  Cryptography (TQC 2019)}}},\ \bibinfo {series} {Leibniz International
  Proceedings in Informatics (LIPIcs)}, Vol.\ \bibinfo {volume} {135}\
  (\bibinfo  {publisher} {Schloss Dagstuhl--Leibniz-Zentrum fuer Informatik},\
  \bibinfo {year} {2019})\ pp.\ \bibinfo {pages} {5:1--5:32}\BibitemShut
  {NoStop}%
\bibitem [{\citenamefont {Kempe}\ \emph {et~al.}(2005)\citenamefont {Kempe},
  \citenamefont {Kitaev},\ and\ \citenamefont {Regev}}]{Kempe2005}%
  \BibitemOpen
  \bibfield  {author} {\bibinfo {author} {\bibfnamefont {J.}~\bibnamefont
  {Kempe}}, \bibinfo {author} {\bibfnamefont {A.}~\bibnamefont {Kitaev}},\ and\
  \bibinfo {author} {\bibfnamefont {O.}~\bibnamefont {Regev}},\ }\bibfield
  {title} {\bibinfo {title} {The complexity of the local hamiltonian problem},\
  }in\ \href {https://doi.org/10.1007/978-3-540-30538-5_31} {\emph {\bibinfo
  {booktitle} {FSTTCS 2004: Foundations of Software Technology and Theoretical
  Computer Science}}}\ (\bibinfo  {publisher} {Springer Berlin Heidelberg},\
  \bibinfo {year} {2005})\ pp.\ \bibinfo {pages} {372--383}\BibitemShut
  {NoStop}%
\bibitem [{\citenamefont {Babbush}\ \emph {et~al.}(2013)\citenamefont
  {Babbush}, \citenamefont {O'Gorman},\ and\ \citenamefont
  {Aspuru-Guzik}}]{Babbush2013}%
  \BibitemOpen
  \bibfield  {author} {\bibinfo {author} {\bibfnamefont {R.}~\bibnamefont
  {Babbush}}, \bibinfo {author} {\bibfnamefont {B.}~\bibnamefont {O'Gorman}},\
  and\ \bibinfo {author} {\bibfnamefont {A.}~\bibnamefont {Aspuru-Guzik}},\
  }\bibfield  {title} {\bibinfo {title} {Resource efficient gadgets for
  compiling adiabatic quantum optimization problems},\ }\href
  {https://doi.org/10.1002/andp.201300120} {\bibfield  {journal} {\bibinfo
  {journal} {Annalen der Physik}\ }\textbf {\bibinfo {volume} {525}},\ \bibinfo
  {pages} {877} (\bibinfo {year} {2013})}\BibitemShut {NoStop}%
\bibitem [{\citenamefont {Zhang}\ \emph {et~al.}(2021)\citenamefont {Zhang},
  \citenamefont {Li}, \citenamefont {Zhang}, \citenamefont {Yuan},
  \citenamefont {Chen}, \citenamefont {Ren}, \citenamefont {Wang},
  \citenamefont {Song}, \citenamefont {Wang}, \citenamefont {Wang} \emph
  {et~al.}}]{Zhang2021}%
  \BibitemOpen
  \bibfield  {author} {\bibinfo {author} {\bibfnamefont {K.}~\bibnamefont
  {Zhang}}, \bibinfo {author} {\bibfnamefont {H.}~\bibnamefont {Li}}, \bibinfo
  {author} {\bibfnamefont {P.}~\bibnamefont {Zhang}}, \bibinfo {author}
  {\bibfnamefont {J.}~\bibnamefont {Yuan}}, \bibinfo {author} {\bibfnamefont
  {J.}~\bibnamefont {Chen}}, \bibinfo {author} {\bibfnamefont {W.}~\bibnamefont
  {Ren}}, \bibinfo {author} {\bibfnamefont {Z.}~\bibnamefont {Wang}}, \bibinfo
  {author} {\bibfnamefont {C.}~\bibnamefont {Song}}, \bibinfo {author}
  {\bibfnamefont {D.-W.}\ \bibnamefont {Wang}}, \bibinfo {author}
  {\bibfnamefont {H.}~\bibnamefont {Wang}}, \emph {et~al.},\ }\bibfield
  {title} {\bibinfo {title} {Synthesizing five-body interaction in a
  superconducting quantum circuit},\ }\href {https://arxiv.org/abs/2109.00964}
  {\bibfield  {journal} {\bibinfo  {journal} {arXiv:2109.00964}\ } (\bibinfo
  {year} {2021})}\BibitemShut {NoStop}%
\bibitem [{\citenamefont {Kranzl}\ \emph {et~al.}(2021)\citenamefont {Kranzl},
  \citenamefont {Joshi}, \citenamefont {Maier}, \citenamefont {Brydges},
  \citenamefont {Franke}, \citenamefont {Blatt},\ and\ \citenamefont
  {Roos}}]{Kranzl2021}%
  \BibitemOpen
  \bibfield  {author} {\bibinfo {author} {\bibfnamefont {F.}~\bibnamefont
  {Kranzl}}, \bibinfo {author} {\bibfnamefont {M.~K.}\ \bibnamefont {Joshi}},
  \bibinfo {author} {\bibfnamefont {C.}~\bibnamefont {Maier}}, \bibinfo
  {author} {\bibfnamefont {T.}~\bibnamefont {Brydges}}, \bibinfo {author}
  {\bibfnamefont {J.}~\bibnamefont {Franke}}, \bibinfo {author} {\bibfnamefont
  {R.}~\bibnamefont {Blatt}},\ and\ \bibinfo {author} {\bibfnamefont {C.~F.}\
  \bibnamefont {Roos}},\ }\bibfield  {title} {\bibinfo {title} {Controlling
  long ion strings for quantum simulation and precision measurements},\ }\href
  {https://arxiv.org/abs/2112.10655} {\bibfield  {journal} {\bibinfo  {journal}
  {arXiv:2112.10655}\ } (\bibinfo {year} {2021})}\BibitemShut {NoStop}%
\bibitem [{\citenamefont {Gu}\ \emph {et~al.}(2021)\citenamefont {Gu},
  \citenamefont {Fern\'andez-Pend\'as}, \citenamefont {Vikst\aa{}l},
  \citenamefont {Abad}, \citenamefont {Warren}, \citenamefont {Bengtsson},
  \citenamefont {Tancredi}, \citenamefont {Shumeiko}, \citenamefont {Bylander},
  \citenamefont {Johansson},\ and\ \citenamefont {Kockum}}]{Gu2021}%
  \BibitemOpen
  \bibfield  {author} {\bibinfo {author} {\bibfnamefont {X.}~\bibnamefont
  {Gu}}, \bibinfo {author} {\bibfnamefont {J.}~\bibnamefont
  {Fern\'andez-Pend\'as}}, \bibinfo {author} {\bibfnamefont {P.}~\bibnamefont
  {Vikst\aa{}l}}, \bibinfo {author} {\bibfnamefont {T.}~\bibnamefont {Abad}},
  \bibinfo {author} {\bibfnamefont {C.}~\bibnamefont {Warren}}, \bibinfo
  {author} {\bibfnamefont {A.}~\bibnamefont {Bengtsson}}, \bibinfo {author}
  {\bibfnamefont {G.}~\bibnamefont {Tancredi}}, \bibinfo {author}
  {\bibfnamefont {V.}~\bibnamefont {Shumeiko}}, \bibinfo {author}
  {\bibfnamefont {J.}~\bibnamefont {Bylander}}, \bibinfo {author}
  {\bibfnamefont {G.}~\bibnamefont {Johansson}},\ and\ \bibinfo {author}
  {\bibfnamefont {A.~F.}\ \bibnamefont {Kockum}},\ }\bibfield  {title}
  {\bibinfo {title} {Fast multiqubit gates through simultaneous two-qubit
  gates},\ }\href {https://doi.org/10.1103/PRXQuantum.2.040348} {\bibfield
  {journal} {\bibinfo  {journal} {PRX Quantum}\ }\textbf {\bibinfo {volume}
  {2}},\ \bibinfo {pages} {040348} (\bibinfo {year} {2021})}\BibitemShut
  {NoStop}%
\bibitem [{\citenamefont {Burkhart}\ \emph {et~al.}(2021)\citenamefont
  {Burkhart}, \citenamefont {Teoh}, \citenamefont {Zhang}, \citenamefont
  {Axline}, \citenamefont {Frunzio}, \citenamefont {Devoret}, \citenamefont
  {Jiang}, \citenamefont {Girvin},\ and\ \citenamefont
  {Schoelkopf}}]{Burkhart2021}%
  \BibitemOpen
  \bibfield  {author} {\bibinfo {author} {\bibfnamefont {L.~D.}\ \bibnamefont
  {Burkhart}}, \bibinfo {author} {\bibfnamefont {J.~D.}\ \bibnamefont {Teoh}},
  \bibinfo {author} {\bibfnamefont {Y.}~\bibnamefont {Zhang}}, \bibinfo
  {author} {\bibfnamefont {C.~J.}\ \bibnamefont {Axline}}, \bibinfo {author}
  {\bibfnamefont {L.}~\bibnamefont {Frunzio}}, \bibinfo {author} {\bibfnamefont
  {M.}~\bibnamefont {Devoret}}, \bibinfo {author} {\bibfnamefont
  {L.}~\bibnamefont {Jiang}}, \bibinfo {author} {\bibfnamefont
  {S.}~\bibnamefont {Girvin}},\ and\ \bibinfo {author} {\bibfnamefont
  {R.}~\bibnamefont {Schoelkopf}},\ }\bibfield  {title} {\bibinfo {title}
  {Error-detected state transfer and entanglement in a superconducting quantum
  network},\ }\href {DOI:https://doi.org/10.1103/PRXQuantum.2.030321}
  {\bibfield  {journal} {\bibinfo  {journal} {PRX Quantum}\ }\textbf {\bibinfo
  {volume} {2}},\ \bibinfo {pages} {030321} (\bibinfo {year}
  {2021})}\BibitemShut {NoStop}%
\bibitem [{\citenamefont {Glaser}\ \emph {et~al.}()\citenamefont {Glaser},
  \citenamefont {Roy},\ and\ \citenamefont {Filipp}}]{Glaser2022}%
  \BibitemOpen
  \bibfield  {author} {\bibinfo {author} {\bibfnamefont {N.}~\bibnamefont
  {Glaser}}, \bibinfo {author} {\bibfnamefont {F.}~\bibnamefont {Roy}},\ and\
  \bibinfo {author} {\bibfnamefont {S.}~\bibnamefont {Filipp}},\ }\bibfield
  {title} {\bibinfo {title} {Engineering a controlled-controlled-phase gate by
  an effective three-body coupling},\ }\href@noop {} {\bibinfo  {journal} {in
  preparation}\ }\BibitemShut {NoStop}%
\bibitem [{\citenamefont {Christandl}\ \emph {et~al.}(2004)\citenamefont
  {Christandl}, \citenamefont {Datta}, \citenamefont {Ekert},\ and\
  \citenamefont {Landahl}}]{Christandl2004}%
  \BibitemOpen
\bibfield  {journal} {  }\bibfield  {author} {\bibinfo {author} {\bibfnamefont
  {M.}~\bibnamefont {Christandl}}, \bibinfo {author} {\bibfnamefont
  {N.}~\bibnamefont {Datta}}, \bibinfo {author} {\bibfnamefont
  {A.}~\bibnamefont {Ekert}},\ and\ \bibinfo {author} {\bibfnamefont {A.~J.}\
  \bibnamefont {Landahl}},\ }\bibfield  {title} {\bibinfo {title} {Perfect
  state transfer in quantum spin networks},\ }\href
  {https://doi.org/10.1103/PhysRevLett.92.187902} {\bibfield  {journal}
  {\bibinfo  {journal} {Phys. Rev. Lett.}\ }\textbf {\bibinfo {volume} {92}},\
  \bibinfo {pages} {187902} (\bibinfo {year} {2004})}\BibitemShut {NoStop}%
\bibitem [{\citenamefont {Yung}(2006)}]{Yung2006}%
  \BibitemOpen
  \bibfield  {author} {\bibinfo {author} {\bibfnamefont {M.-H.}\ \bibnamefont
  {Yung}},\ }\bibfield  {title} {\bibinfo {title} {Quantum speed limit for
  perfect state transfer in one dimension},\ }\href
  {https://doi.org/10.1103/PhysRevA.74.030303} {\bibfield  {journal} {\bibinfo
  {journal} {Phys. Rev. A}\ }\textbf {\bibinfo {volume} {74}},\ \bibinfo
  {pages} {030303} (\bibinfo {year} {2006})}\BibitemShut {NoStop}%
\bibitem [{\citenamefont {Vinet}\ and\ \citenamefont
  {Zhedanov}(2012{\natexlab{a}})}]{Vinet2012}%
  \BibitemOpen
  \bibfield  {author} {\bibinfo {author} {\bibfnamefont {L.}~\bibnamefont
  {Vinet}}\ and\ \bibinfo {author} {\bibfnamefont {A.}~\bibnamefont
  {Zhedanov}},\ }\bibfield  {title} {\bibinfo {title} {How to construct spin
  chains with perfect state transfer},\ }\href
  {https://doi.org/10.1103/PhysRevA.85.012323} {\bibfield  {journal} {\bibinfo
  {journal} {Phys. Rev. A}\ }\textbf {\bibinfo {volume} {85}},\ \bibinfo
  {pages} {012323} (\bibinfo {year} {2012}{\natexlab{a}})}\BibitemShut
  {NoStop}%
\bibitem [{\citenamefont {Chapman}\ \emph {et~al.}(2016)\citenamefont
  {Chapman}, \citenamefont {Santandrea}, \citenamefont {Huang}, \citenamefont
  {Corrielli}, \citenamefont {Crespi}, \citenamefont {Yung}, \citenamefont
  {Osellame},\ and\ \citenamefont {Peruzzo}}]{Chapman2016}%
  \BibitemOpen
  \bibfield  {author} {\bibinfo {author} {\bibfnamefont {R.~J.}\ \bibnamefont
  {Chapman}}, \bibinfo {author} {\bibfnamefont {M.}~\bibnamefont {Santandrea}},
  \bibinfo {author} {\bibfnamefont {Z.}~\bibnamefont {Huang}}, \bibinfo
  {author} {\bibfnamefont {G.}~\bibnamefont {Corrielli}}, \bibinfo {author}
  {\bibfnamefont {A.}~\bibnamefont {Crespi}}, \bibinfo {author} {\bibfnamefont
  {M.-H.}\ \bibnamefont {Yung}}, \bibinfo {author} {\bibfnamefont
  {R.}~\bibnamefont {Osellame}},\ and\ \bibinfo {author} {\bibfnamefont
  {A.}~\bibnamefont {Peruzzo}},\ }\bibfield  {title} {\bibinfo {title}
  {Experimental perfect state transfer of an entangled photonic qubit},\ }\href
  {https://doi.org/10.1038/ncomms11339} {\bibfield  {journal} {\bibinfo
  {journal} {Nature communications}\ }\textbf {\bibinfo {volume} {7}},\
  \bibinfo {pages} {1} (\bibinfo {year} {2016})}\BibitemShut {NoStop}%
\bibitem [{\citenamefont {Brougham}\ \emph {et~al.}(2011)\citenamefont
  {Brougham}, \citenamefont {Nikolopoulos},\ and\ \citenamefont
  {Jex}}]{Brougham2011}%
  \BibitemOpen
  \bibfield  {author} {\bibinfo {author} {\bibfnamefont {T.}~\bibnamefont
  {Brougham}}, \bibinfo {author} {\bibfnamefont {G.}~\bibnamefont
  {Nikolopoulos}},\ and\ \bibinfo {author} {\bibfnamefont {I.}~\bibnamefont
  {Jex}},\ }\bibfield  {title} {\bibinfo {title} {Perfect transfer of multiple
  excitations in quantum networks},\ }\href
  {https://doi.org/10.1103/PhysRevA.83.022323} {\bibfield  {journal} {\bibinfo
  {journal} {Phys. Rev. A}\ }\textbf {\bibinfo {volume} {83}},\ \bibinfo
  {pages} {022323} (\bibinfo {year} {2011})}\BibitemShut {NoStop}%
\bibitem [{\citenamefont {Nikolopoulos}\ \emph {et~al.}(2004)\citenamefont
  {Nikolopoulos}, \citenamefont {Petrosyan},\ and\ \citenamefont
  {Lambropoulos}}]{Nikolopoulos2004}%
  \BibitemOpen
  \bibfield  {author} {\bibinfo {author} {\bibfnamefont {G.~M.}\ \bibnamefont
  {Nikolopoulos}}, \bibinfo {author} {\bibfnamefont {D.}~\bibnamefont
  {Petrosyan}},\ and\ \bibinfo {author} {\bibfnamefont {P.}~\bibnamefont
  {Lambropoulos}},\ }\bibfield  {title} {\bibinfo {title} {Electron wavepacket
  propagation in a chain of coupled quantum dots},\ }\href
  {https://doi.org/10.1088/0953-8984/16/28/019} {\bibfield  {journal} {\bibinfo
   {journal} {Journal of Physics: Condensed Matter}\ }\textbf {\bibinfo
  {volume} {16}},\ \bibinfo {pages} {4991} (\bibinfo {year}
  {2004})}\BibitemShut {NoStop}%
\bibitem [{\citenamefont {Shi}\ \emph {et~al.}(2005)\citenamefont {Shi},
  \citenamefont {Li}, \citenamefont {Song},\ and\ \citenamefont
  {Sun}}]{Shi2005}%
  \BibitemOpen
  \bibfield  {author} {\bibinfo {author} {\bibfnamefont {T.}~\bibnamefont
  {Shi}}, \bibinfo {author} {\bibfnamefont {Y.}~\bibnamefont {Li}}, \bibinfo
  {author} {\bibfnamefont {Z.}~\bibnamefont {Song}},\ and\ \bibinfo {author}
  {\bibfnamefont {C.-P.}\ \bibnamefont {Sun}},\ }\bibfield  {title} {\bibinfo
  {title} {Quantum-state transfer via the ferromagnetic chain in a spatially
  modulated field},\ }\href {https://doi.org/10.1103/PhysRevA.71.032309}
  {\bibfield  {journal} {\bibinfo  {journal} {Phys. Rev. A}\ }\textbf {\bibinfo
  {volume} {71}},\ \bibinfo {pages} {032309} (\bibinfo {year}
  {2005})}\BibitemShut {NoStop}%
\bibitem [{\citenamefont {Albanese}\ \emph {et~al.}(2004)\citenamefont
  {Albanese}, \citenamefont {Christandl}, \citenamefont {Datta},\ and\
  \citenamefont {Ekert}}]{Albanese2004}%
  \BibitemOpen
  \bibfield  {author} {\bibinfo {author} {\bibfnamefont {C.}~\bibnamefont
  {Albanese}}, \bibinfo {author} {\bibfnamefont {M.}~\bibnamefont
  {Christandl}}, \bibinfo {author} {\bibfnamefont {N.}~\bibnamefont {Datta}},\
  and\ \bibinfo {author} {\bibfnamefont {A.}~\bibnamefont {Ekert}},\ }\bibfield
   {title} {\bibinfo {title} {Mirror inversion of quantum states in linear
  registers},\ }\href {https://doi.org/10.1103/PhysRevLett.93.230502}
  {\bibfield  {journal} {\bibinfo  {journal} {Phys. Rev. Lett.}\ }\textbf
  {\bibinfo {volume} {93}},\ \bibinfo {pages} {230502} (\bibinfo {year}
  {2004})}\BibitemShut {NoStop}%
\bibitem [{\citenamefont {Vinet}\ and\ \citenamefont
  {Zhedanov}(2012{\natexlab{b}})}]{Vinet2012_Krawtchouk}%
  \BibitemOpen
  \bibfield  {author} {\bibinfo {author} {\bibfnamefont {L.}~\bibnamefont
  {Vinet}}\ and\ \bibinfo {author} {\bibfnamefont {A.}~\bibnamefont
  {Zhedanov}},\ }\bibfield  {title} {\bibinfo {title} {Para-krawtchouk
  polynomials on a bi-lattice and a quantum spin chain with perfect state
  transfer},\ }\href {https://doi.org/10.1088/1751-8113/45/26/265304}
  {\bibfield  {journal} {\bibinfo  {journal} {Journal of Physics A:
  Mathematical and Theoretical}\ }\textbf {\bibinfo {volume} {45}},\ \bibinfo
  {pages} {265304} (\bibinfo {year} {2012}{\natexlab{b}})}\BibitemShut
  {NoStop}%
\bibitem [{\citenamefont {Yung}\ and\ \citenamefont {Bose}(2005)}]{Yung2005}%
  \BibitemOpen
  \bibfield  {author} {\bibinfo {author} {\bibfnamefont {M.-H.}\ \bibnamefont
  {Yung}}\ and\ \bibinfo {author} {\bibfnamefont {S.}~\bibnamefont {Bose}},\
  }\bibfield  {title} {\bibinfo {title} {Perfect state transfer, effective
  gates, and entanglement generation in engineered bosonic and fermionic
  networks},\ }\href {https://doi.org/10.1103/PhysRevA.71.032310} {\bibfield
  {journal} {\bibinfo  {journal} {Phys. Rev. A}\ }\textbf {\bibinfo {volume}
  {71}},\ \bibinfo {pages} {032310} (\bibinfo {year} {2005})}\BibitemShut
  {NoStop}%
\bibitem [{\citenamefont {Kay}(2010)}]{Kay2010}%
  \BibitemOpen
  \bibfield  {author} {\bibinfo {author} {\bibfnamefont {A.}~\bibnamefont
  {Kay}},\ }\bibfield  {title} {\bibinfo {title} {Perfect, efficient, state
  transfer and its application as a constructive tool},\ }\href
  {https://doi.org/10.1142/S0219749910006514} {\bibfield  {journal} {\bibinfo
  {journal} {International Journal of Quantum Information}\ }\textbf {\bibinfo
  {volume} {8}},\ \bibinfo {pages} {641} (\bibinfo {year} {2010})}\BibitemShut
  {NoStop}%
\bibitem [{\citenamefont {Li}\ \emph {et~al.}(2018)\citenamefont {Li},
  \citenamefont {Ma}, \citenamefont {Han}, \citenamefont {Chen}, \citenamefont
  {Xu}, \citenamefont {Cai}, \citenamefont {Wang}, \citenamefont {Song},
  \citenamefont {Xue}, \citenamefont {Yin} \emph {et~al.}}]{Li2018}%
  \BibitemOpen
  \bibfield  {author} {\bibinfo {author} {\bibfnamefont {X.}~\bibnamefont
  {Li}}, \bibinfo {author} {\bibfnamefont {Y.}~\bibnamefont {Ma}}, \bibinfo
  {author} {\bibfnamefont {J.}~\bibnamefont {Han}}, \bibinfo {author}
  {\bibfnamefont {T.}~\bibnamefont {Chen}}, \bibinfo {author} {\bibfnamefont
  {Y.}~\bibnamefont {Xu}}, \bibinfo {author} {\bibfnamefont {W.}~\bibnamefont
  {Cai}}, \bibinfo {author} {\bibfnamefont {H.}~\bibnamefont {Wang}}, \bibinfo
  {author} {\bibfnamefont {Y.}~\bibnamefont {Song}}, \bibinfo {author}
  {\bibfnamefont {Z.-Y.}\ \bibnamefont {Xue}}, \bibinfo {author} {\bibfnamefont
  {Z.-q.}\ \bibnamefont {Yin}}, \emph {et~al.},\ }\bibfield  {title} {\bibinfo
  {title} {Perfect quantum state transfer in a superconducting qubit chain with
  parametrically tunable couplings},\ }\href
  {https://doi.org/10.1103/PhysRevApplied.10.054009} {\bibfield  {journal}
  {\bibinfo  {journal} {Phys. Rev. Applied}\ }\textbf {\bibinfo {volume}
  {10}},\ \bibinfo {pages} {054009} (\bibinfo {year} {2018})}\BibitemShut
  {NoStop}%
\bibitem [{\citenamefont {Nielsen}\ and\ \citenamefont
  {Chuang}(2010)}]{Nielsen2002}%
  \BibitemOpen
  \bibfield  {author} {\bibinfo {author} {\bibfnamefont {M.~A.}\ \bibnamefont
  {Nielsen}}\ and\ \bibinfo {author} {\bibfnamefont {I.~L.}\ \bibnamefont
  {Chuang}},\ }\href {https://doi.org/10.1017/CBO9780511976667} {\emph
  {\bibinfo {title} {Quantum Computation and Quantum Information: 10th
  Anniversary Edition}}}\ (\bibinfo  {publisher} {Cambridge University Press},\
  \bibinfo {year} {2010})\BibitemShut {NoStop}%
\bibitem [{\citenamefont {Yung}\ \emph {et~al.}(2004)\citenamefont {Yung},
  \citenamefont {Leung},\ and\ \citenamefont {Bose}}]{Yung2004}%
  \BibitemOpen
  \bibfield  {author} {\bibinfo {author} {\bibfnamefont {M.~H.}\ \bibnamefont
  {Yung}}, \bibinfo {author} {\bibfnamefont {D.}~\bibnamefont {Leung}},\ and\
  \bibinfo {author} {\bibfnamefont {S.}~\bibnamefont {Bose}},\ }\bibfield
  {title} {\bibinfo {title} {An exact effective two-qubit gate in a chain of
  three spins},\ }\href {https://doi.org/10.26421/QIC4.3-2} {\bibfield
  {journal} {\bibinfo  {journal} {Quantum Information and Computation}\
  }\textbf {\bibinfo {volume} {4}} (\bibinfo {year} {2004})}\BibitemShut
  {NoStop}%
\bibitem [{\citenamefont {Genest}\ \emph {et~al.}(2016)\citenamefont {Genest},
  \citenamefont {Vinet},\ and\ \citenamefont {Zhedanov}}]{Genest2016}%
  \BibitemOpen
  \bibfield  {author} {\bibinfo {author} {\bibfnamefont {V.~X.}\ \bibnamefont
  {Genest}}, \bibinfo {author} {\bibfnamefont {L.}~\bibnamefont {Vinet}},\ and\
  \bibinfo {author} {\bibfnamefont {A.}~\bibnamefont {Zhedanov}},\ }\bibfield
  {title} {\bibinfo {title} {Quantum spin chains with fractional revival},\
  }\href {https://doi.org/10.1016/j.aop.2016.05.009} {\bibfield  {journal}
  {\bibinfo  {journal} {Annals of Physics}\ }\textbf {\bibinfo {volume}
  {371}},\ \bibinfo {pages} {348} (\bibinfo {year} {2016})}\BibitemShut
  {NoStop}%
\bibitem [{\citenamefont {Chan}\ \emph {et~al.}(2019)\citenamefont {Chan},
  \citenamefont {Coutinho}, \citenamefont {Tamon}, \citenamefont {Vinet},\ and\
  \citenamefont {Zhan}}]{Chan2019}%
  \BibitemOpen
  \bibfield  {author} {\bibinfo {author} {\bibfnamefont {A.}~\bibnamefont
  {Chan}}, \bibinfo {author} {\bibfnamefont {G.}~\bibnamefont {Coutinho}},
  \bibinfo {author} {\bibfnamefont {C.}~\bibnamefont {Tamon}}, \bibinfo
  {author} {\bibfnamefont {L.}~\bibnamefont {Vinet}},\ and\ \bibinfo {author}
  {\bibfnamefont {H.}~\bibnamefont {Zhan}},\ }\bibfield  {title} {\bibinfo
  {title} {Quantum fractional revival on graphs},\ }\href
  {https://doi.org/10.1016/j.dam.2018.12.017} {\bibfield  {journal} {\bibinfo
  {journal} {Discrete Applied Mathematics}\ }\textbf {\bibinfo {volume}
  {269}},\ \bibinfo {pages} {86} (\bibinfo {year} {2019})}\BibitemShut
  {NoStop}%
\bibitem [{\citenamefont {Lemay}\ \emph {et~al.}(2016)\citenamefont {Lemay},
  \citenamefont {Vinet},\ and\ \citenamefont {Zhedanov}}]{Lemay2016}%
  \BibitemOpen
  \bibfield  {author} {\bibinfo {author} {\bibfnamefont {J.-M.}\ \bibnamefont
  {Lemay}}, \bibinfo {author} {\bibfnamefont {L.}~\bibnamefont {Vinet}},\ and\
  \bibinfo {author} {\bibfnamefont {A.}~\bibnamefont {Zhedanov}},\ }\bibfield
  {title} {\bibinfo {title} {An analytic spin chain model with fractional
  revival},\ }\href {https://doi.org/10.1088/1751-8113/49/33/335302} {\bibfield
   {journal} {\bibinfo  {journal} {Journal of Physics A: Mathematical and
  Theoretical}\ }\textbf {\bibinfo {volume} {49}},\ \bibinfo {pages} {335302}
  (\bibinfo {year} {2016})}\BibitemShut {NoStop}%
\bibitem [{\citenamefont {Valiant}(2002)}]{Valiant2002}%
  \BibitemOpen
  \bibfield  {author} {\bibinfo {author} {\bibfnamefont {L.~G.}\ \bibnamefont
  {Valiant}},\ }\bibfield  {title} {\bibinfo {title} {Quantum circuits that can
  be simulated classically in polynomial time},\ }\href
  {https://doi.org/10.1137/S0097539700377025} {\bibfield  {journal} {\bibinfo
  {journal} {SIAM Journal on Computing}\ }\textbf {\bibinfo {volume} {31}},\
  \bibinfo {pages} {1229} (\bibinfo {year} {2002})}\BibitemShut {NoStop}%
\bibitem [{\citenamefont {Terhal}\ and\ \citenamefont
  {DiVincenzo}(2002)}]{Terhal2002}%
  \BibitemOpen
  \bibfield  {author} {\bibinfo {author} {\bibfnamefont {B.~M.}\ \bibnamefont
  {Terhal}}\ and\ \bibinfo {author} {\bibfnamefont {D.~P.}\ \bibnamefont
  {DiVincenzo}},\ }\bibfield  {title} {\bibinfo {title} {Classical simulation
  of noninteracting-fermion quantum circuits},\ }\href
  {https://doi.org/10.1103/PhysRevA.65.032325} {\bibfield  {journal} {\bibinfo
  {journal} {Phys. Rev. A}\ }\textbf {\bibinfo {volume} {65}},\ \bibinfo
  {pages} {032325} (\bibinfo {year} {2002})}\BibitemShut {NoStop}%
\bibitem [{\citenamefont {Knill}(2001)}]{Knill2001}%
  \BibitemOpen
  \bibfield  {author} {\bibinfo {author} {\bibfnamefont {E.}~\bibnamefont
  {Knill}},\ }\bibfield  {title} {\bibinfo {title} {Fermionic linear optics and
  matchgates},\ }\href {https://arxiv.org/abs/quant-ph/0108033} {\bibfield
  {journal} {\bibinfo  {journal} {arXiv:quant-ph/0108033}\ } (\bibinfo {year}
  {2001})}\BibitemShut {NoStop}%
\bibitem [{\citenamefont {Cohen}\ \emph {et~al.}(2021)\citenamefont {Cohen},
  \citenamefont {Kim}, \citenamefont {Bartlett},\ and\ \citenamefont
  {Brown}}]{Cohen2021}%
  \BibitemOpen
  \bibfield  {author} {\bibinfo {author} {\bibfnamefont {L.~Z.}\ \bibnamefont
  {Cohen}}, \bibinfo {author} {\bibfnamefont {I.~H.}\ \bibnamefont {Kim}},
  \bibinfo {author} {\bibfnamefont {S.~D.}\ \bibnamefont {Bartlett}},\ and\
  \bibinfo {author} {\bibfnamefont {B.~J.}\ \bibnamefont {Brown}},\ }\bibfield
  {title} {\bibinfo {title} {Low-overhead fault-tolerant quantum computing
  using long-range connectivity},\ }\href {https://arxiv.org/abs/2110.10794}
  {\bibfield  {journal} {\bibinfo  {journal} {arXiv:2110.10794}\ } (\bibinfo
  {year} {2021})}\BibitemShut {NoStop}%
\bibitem [{\citenamefont {Lidar}\ and\ \citenamefont {Brun}(2013)}]{Lidar2013}%
  \BibitemOpen
  \bibfield  {author} {\bibinfo {author} {\bibfnamefont {D.}~\bibnamefont
  {Lidar}}\ and\ \bibinfo {author} {\bibfnamefont {T.}~\bibnamefont {Brun}},\
  }\href {https://doi.org/10.1017/CBO9781139034807} {\emph {\bibinfo {title}
  {Quantum Error Correction}}}\ (\bibinfo  {publisher} {Cambridge University
  Press},\ \bibinfo {year} {2013})\BibitemShut {NoStop}%
\bibitem [{\citenamefont {Hochstadt}(1967)}]{Hochstadt1967}%
  \BibitemOpen
  \bibfield  {author} {\bibinfo {author} {\bibfnamefont {H.}~\bibnamefont
  {Hochstadt}},\ }\bibfield  {title} {\bibinfo {title} {On some inverse
  problems in matrix theory},\ }\href {https://doi.org/10.1007/BF01899647}
  {\bibfield  {journal} {\bibinfo  {journal} {Archiv der Mathematik}\ }\textbf
  {\bibinfo {volume} {18}},\ \bibinfo {pages} {201} (\bibinfo {year}
  {1967})}\BibitemShut {NoStop}%
\bibitem [{\citenamefont {Cappellaro}\ \emph {et~al.}(2011)\citenamefont
  {Cappellaro}, \citenamefont {Viola},\ and\ \citenamefont
  {Ramanathan}}]{Cappellaro2011}%
  \BibitemOpen
  \bibfield  {author} {\bibinfo {author} {\bibfnamefont {P.}~\bibnamefont
  {Cappellaro}}, \bibinfo {author} {\bibfnamefont {L.}~\bibnamefont {Viola}},\
  and\ \bibinfo {author} {\bibfnamefont {C.}~\bibnamefont {Ramanathan}},\
  }\bibfield  {title} {\bibinfo {title} {Coherent-state transfer via highly
  mixed quantum spin chains},\ }\href
  {https://doi.org/10.1103/PhysRevA.83.032304} {\bibfield  {journal} {\bibinfo
  {journal} {Phys. Rev. A}\ }\textbf {\bibinfo {volume} {83}},\ \bibinfo
  {pages} {032304} (\bibinfo {year} {2011})}\BibitemShut {NoStop}%
\bibitem [{\citenamefont {Ortiz}\ \emph {et~al.}(2001)\citenamefont {Ortiz},
  \citenamefont {Gubernatis}, \citenamefont {Knill},\ and\ \citenamefont
  {Laflamme}}]{Ortiz2001}%
  \BibitemOpen
  \bibfield  {author} {\bibinfo {author} {\bibfnamefont {G.}~\bibnamefont
  {Ortiz}}, \bibinfo {author} {\bibfnamefont {J.~E.}\ \bibnamefont
  {Gubernatis}}, \bibinfo {author} {\bibfnamefont {E.}~\bibnamefont {Knill}},\
  and\ \bibinfo {author} {\bibfnamefont {R.}~\bibnamefont {Laflamme}},\
  }\bibfield  {title} {\bibinfo {title} {Quantum algorithms for fermionic
  simulations},\ }\href {https://doi.org/10.1103/PhysRevA.64.022319} {\bibfield
   {journal} {\bibinfo  {journal} {Phys. Rev. A}\ }\textbf {\bibinfo {volume}
  {64}},\ \bibinfo {pages} {022319} (\bibinfo {year} {2001})}\BibitemShut
  {NoStop}%
\bibitem [{\citenamefont {Cade}\ \emph {et~al.}(2020)\citenamefont {Cade},
  \citenamefont {Mineh}, \citenamefont {Montanaro},\ and\ \citenamefont
  {Stanisic}}]{Cade2020}%
  \BibitemOpen
  \bibfield  {author} {\bibinfo {author} {\bibfnamefont {C.}~\bibnamefont
  {Cade}}, \bibinfo {author} {\bibfnamefont {L.}~\bibnamefont {Mineh}},
  \bibinfo {author} {\bibfnamefont {A.}~\bibnamefont {Montanaro}},\ and\
  \bibinfo {author} {\bibfnamefont {S.}~\bibnamefont {Stanisic}},\ }\bibfield
  {title} {\bibinfo {title} {Strategies for solving the fermi-hubbard model on
  near-term quantum computers},\ }\href
  {https://doi.org/10.1103/PhysRevB.102.235122} {\bibfield  {journal} {\bibinfo
   {journal} {Phys. Rev. B}\ }\textbf {\bibinfo {volume} {102}},\ \bibinfo
  {pages} {235122} (\bibinfo {year} {2020})}\BibitemShut {NoStop}%
\bibitem [{\citenamefont {Lieb}\ \emph {et~al.}(1961)\citenamefont {Lieb},
  \citenamefont {Schultz},\ and\ \citenamefont {Mattis}}]{Lieb1961}%
  \BibitemOpen
  \bibfield  {author} {\bibinfo {author} {\bibfnamefont {E.}~\bibnamefont
  {Lieb}}, \bibinfo {author} {\bibfnamefont {T.}~\bibnamefont {Schultz}},\ and\
  \bibinfo {author} {\bibfnamefont {D.}~\bibnamefont {Mattis}},\ }\bibfield
  {title} {\bibinfo {title} {Two soluble models of an antiferromagnetic
  chain},\ }\href {https://doi.org/10.1016/0003-4916(61)90115-4} {\bibfield
  {journal} {\bibinfo  {journal} {Annals of Physics}\ }\textbf {\bibinfo
  {volume} {16}},\ \bibinfo {pages} {407} (\bibinfo {year} {1961})}\BibitemShut
  {NoStop}%
\bibitem [{\citenamefont {Borjans}\ \emph {et~al.}(2020)\citenamefont
  {Borjans}, \citenamefont {Croot}, \citenamefont {Mi}, \citenamefont
  {Gullans},\ and\ \citenamefont {Petta}}]{Borjans2020}%
  \BibitemOpen
  \bibfield  {author} {\bibinfo {author} {\bibfnamefont {F.}~\bibnamefont
  {Borjans}}, \bibinfo {author} {\bibfnamefont {X.}~\bibnamefont {Croot}},
  \bibinfo {author} {\bibfnamefont {X.}~\bibnamefont {Mi}}, \bibinfo {author}
  {\bibfnamefont {M.}~\bibnamefont {Gullans}},\ and\ \bibinfo {author}
  {\bibfnamefont {J.}~\bibnamefont {Petta}},\ }\bibfield  {title} {\bibinfo
  {title} {Resonant microwave-mediated interactions between distant electron
  spins},\ }\href {https://doi.org/10.1038/s41586-019-1867-y} {\bibfield
  {journal} {\bibinfo  {journal} {Nature}\ }\textbf {\bibinfo {volume} {577}},\
  \bibinfo {pages} {195} (\bibinfo {year} {2020})}\BibitemShut {NoStop}%
\bibitem [{\citenamefont {Koch}\ \emph {et~al.}(2007)\citenamefont {Koch},
  \citenamefont {Terri}, \citenamefont {Gambetta}, \citenamefont {Houck},
  \citenamefont {Schuster}, \citenamefont {Majer}, \citenamefont {Blais},
  \citenamefont {Devoret}, \citenamefont {Girvin},\ and\ \citenamefont
  {Schoelkopf}}]{Koch2007}%
  \BibitemOpen
  \bibfield  {author} {\bibinfo {author} {\bibfnamefont {J.}~\bibnamefont
  {Koch}}, \bibinfo {author} {\bibfnamefont {M.~Y.}\ \bibnamefont {Terri}},
  \bibinfo {author} {\bibfnamefont {J.}~\bibnamefont {Gambetta}}, \bibinfo
  {author} {\bibfnamefont {A.~A.}\ \bibnamefont {Houck}}, \bibinfo {author}
  {\bibfnamefont {D.~I.}\ \bibnamefont {Schuster}}, \bibinfo {author}
  {\bibfnamefont {J.}~\bibnamefont {Majer}}, \bibinfo {author} {\bibfnamefont
  {A.}~\bibnamefont {Blais}}, \bibinfo {author} {\bibfnamefont {M.~H.}\
  \bibnamefont {Devoret}}, \bibinfo {author} {\bibfnamefont {S.~M.}\
  \bibnamefont {Girvin}},\ and\ \bibinfo {author} {\bibfnamefont {R.~J.}\
  \bibnamefont {Schoelkopf}},\ }\bibfield  {title} {\bibinfo {title}
  {Charge-insensitive qubit design derived from the cooper pair box},\ }\href
  {https://doi.org/10.1103/PhysRevA.76.042319} {\bibfield  {journal} {\bibinfo
  {journal} {Phys. Rev. A}\ }\textbf {\bibinfo {volume} {76}},\ \bibinfo
  {pages} {042319} (\bibinfo {year} {2007})}\BibitemShut {NoStop}%
\bibitem [{\citenamefont {Hutchings}\ \emph {et~al.}(2017)\citenamefont
  {Hutchings}, \citenamefont {Hertzberg}, \citenamefont {Liu}, \citenamefont
  {Bronn}, \citenamefont {Keefe}, \citenamefont {Brink}, \citenamefont {Chow},\
  and\ \citenamefont {Plourde}}]{Hutchings2017}%
  \BibitemOpen
  \bibfield  {author} {\bibinfo {author} {\bibfnamefont {M.~D.}\ \bibnamefont
  {Hutchings}}, \bibinfo {author} {\bibfnamefont {J.~B.}\ \bibnamefont
  {Hertzberg}}, \bibinfo {author} {\bibfnamefont {Y.}~\bibnamefont {Liu}},
  \bibinfo {author} {\bibfnamefont {N.~T.}\ \bibnamefont {Bronn}}, \bibinfo
  {author} {\bibfnamefont {G.~A.}\ \bibnamefont {Keefe}}, \bibinfo {author}
  {\bibfnamefont {M.}~\bibnamefont {Brink}}, \bibinfo {author} {\bibfnamefont
  {J.~M.}\ \bibnamefont {Chow}},\ and\ \bibinfo {author} {\bibfnamefont
  {B.~L.~T.}\ \bibnamefont {Plourde}},\ }\bibfield  {title} {\bibinfo {title}
  {Tunable superconducting qubits with flux-independent coherence},\ }\href
  {https://doi.org/10.1103/PhysRevApplied.8.044003} {\bibfield  {journal}
  {\bibinfo  {journal} {Phys. Rev. Applied}\ }\textbf {\bibinfo {volume} {8}},\
  \bibinfo {pages} {044003} (\bibinfo {year} {2017})}\BibitemShut {NoStop}%
\bibitem [{\citenamefont {McClean}\ \emph {et~al.}(2016)\citenamefont
  {McClean}, \citenamefont {Romero}, \citenamefont {Babbush},\ and\
  \citenamefont {Aspuru-Guzik}}]{Mcclean2016}%
  \BibitemOpen
  \bibfield  {author} {\bibinfo {author} {\bibfnamefont {J.~R.}\ \bibnamefont
  {McClean}}, \bibinfo {author} {\bibfnamefont {J.}~\bibnamefont {Romero}},
  \bibinfo {author} {\bibfnamefont {R.}~\bibnamefont {Babbush}},\ and\ \bibinfo
  {author} {\bibfnamefont {A.}~\bibnamefont {Aspuru-Guzik}},\ }\bibfield
  {title} {\bibinfo {title} {The theory of variational hybrid quantum-classical
  algorithms},\ }\href {https://doi.org/10.1088/1367-2630/18/2/023023}
  {\bibfield  {journal} {\bibinfo  {journal} {New Journal of Physics}\ }\textbf
  {\bibinfo {volume} {18}},\ \bibinfo {pages} {023023} (\bibinfo {year}
  {2016})}\BibitemShut {NoStop}%
\bibitem [{\citenamefont {Wittler}\ \emph {et~al.}(2021)\citenamefont
  {Wittler}, \citenamefont {Roy}, \citenamefont {Pack}, \citenamefont
  {Werninghaus}, \citenamefont {Roy}, \citenamefont {Egger}, \citenamefont
  {Filipp}, \citenamefont {Wilhelm},\ and\ \citenamefont
  {Machnes}}]{Wittler2021}%
  \BibitemOpen
  \bibfield  {author} {\bibinfo {author} {\bibfnamefont {N.}~\bibnamefont
  {Wittler}}, \bibinfo {author} {\bibfnamefont {F.}~\bibnamefont {Roy}},
  \bibinfo {author} {\bibfnamefont {K.}~\bibnamefont {Pack}}, \bibinfo {author}
  {\bibfnamefont {M.}~\bibnamefont {Werninghaus}}, \bibinfo {author}
  {\bibfnamefont {A.~S.}\ \bibnamefont {Roy}}, \bibinfo {author} {\bibfnamefont
  {D.~J.}\ \bibnamefont {Egger}}, \bibinfo {author} {\bibfnamefont
  {S.}~\bibnamefont {Filipp}}, \bibinfo {author} {\bibfnamefont {F.~K.}\
  \bibnamefont {Wilhelm}},\ and\ \bibinfo {author} {\bibfnamefont
  {S.}~\bibnamefont {Machnes}},\ }\bibfield  {title} {\bibinfo {title}
  {Integrated tool set for control, calibration, and characterization of
  quantum devices applied to superconducting qubits},\ }\href
  {https://doi.org/10.1103/PhysRevApplied.15.034080} {\bibfield  {journal}
  {\bibinfo  {journal} {Phys. Rev. Applied}\ }\textbf {\bibinfo {volume}
  {15}},\ \bibinfo {pages} {034080} (\bibinfo {year} {2021})}\BibitemShut
  {NoStop}%
\bibitem [{\citenamefont {McKay}\ \emph {et~al.}(2016)\citenamefont {McKay},
  \citenamefont {Filipp}, \citenamefont {Mezzacapo}, \citenamefont {Magesan},
  \citenamefont {Chow},\ and\ \citenamefont {Gambetta}}]{Mckay2016}%
  \BibitemOpen
  \bibfield  {author} {\bibinfo {author} {\bibfnamefont {D.~C.}\ \bibnamefont
  {McKay}}, \bibinfo {author} {\bibfnamefont {S.}~\bibnamefont {Filipp}},
  \bibinfo {author} {\bibfnamefont {A.}~\bibnamefont {Mezzacapo}}, \bibinfo
  {author} {\bibfnamefont {E.}~\bibnamefont {Magesan}}, \bibinfo {author}
  {\bibfnamefont {J.~M.}\ \bibnamefont {Chow}},\ and\ \bibinfo {author}
  {\bibfnamefont {J.~M.}\ \bibnamefont {Gambetta}},\ }\bibfield  {title}
  {\bibinfo {title} {Universal gate for fixed-frequency qubits via a tunable
  bus},\ }\href {https://doi.org/10.1103/PhysRevApplied.6.064007} {\bibfield
  {journal} {\bibinfo  {journal} {Phys. Rev. Applied}\ }\textbf {\bibinfo
  {volume} {6}},\ \bibinfo {pages} {064007} (\bibinfo {year}
  {2016})}\BibitemShut {NoStop}%
\bibitem [{\citenamefont {Sete}\ \emph {et~al.}(2021)\citenamefont {Sete},
  \citenamefont {Didier}, \citenamefont {Chen}, \citenamefont {Kulshreshtha},
  \citenamefont {Manenti},\ and\ \citenamefont {Poletto}}]{Sete2021}%
  \BibitemOpen
  \bibfield  {author} {\bibinfo {author} {\bibfnamefont {E.~A.}\ \bibnamefont
  {Sete}}, \bibinfo {author} {\bibfnamefont {N.}~\bibnamefont {Didier}},
  \bibinfo {author} {\bibfnamefont {A.~Q.}\ \bibnamefont {Chen}}, \bibinfo
  {author} {\bibfnamefont {S.}~\bibnamefont {Kulshreshtha}}, \bibinfo {author}
  {\bibfnamefont {R.}~\bibnamefont {Manenti}},\ and\ \bibinfo {author}
  {\bibfnamefont {S.}~\bibnamefont {Poletto}},\ }\bibfield  {title} {\bibinfo
  {title} {Parametric-resonance entangling gates with a tunable coupler},\
  }\href {https://doi.org/10.1103/PhysRevApplied.16.024050} {\bibfield
  {journal} {\bibinfo  {journal} {Phys. Rev. Applied}\ }\textbf {\bibinfo
  {volume} {16}},\ \bibinfo {pages} {024050} (\bibinfo {year}
  {2021})}\BibitemShut {NoStop}%
\bibitem [{\citenamefont {McKay}\ \emph {et~al.}(2017)\citenamefont {McKay},
  \citenamefont {Wood}, \citenamefont {Sheldon}, \citenamefont {Chow},\ and\
  \citenamefont {Gambetta}}]{Mckay2017}%
  \BibitemOpen
  \bibfield  {author} {\bibinfo {author} {\bibfnamefont {D.~C.}\ \bibnamefont
  {McKay}}, \bibinfo {author} {\bibfnamefont {C.~J.}\ \bibnamefont {Wood}},
  \bibinfo {author} {\bibfnamefont {S.}~\bibnamefont {Sheldon}}, \bibinfo
  {author} {\bibfnamefont {J.~M.}\ \bibnamefont {Chow}},\ and\ \bibinfo
  {author} {\bibfnamefont {J.~M.}\ \bibnamefont {Gambetta}},\ }\bibfield
  {title} {\bibinfo {title} {Efficient {Z} gates for quantum computing},\
  }\href {https://doi.org/10.1103/PhysRevA.96.022330} {\bibfield  {journal}
  {\bibinfo  {journal} {Phys. Rev. A}\ }\textbf {\bibinfo {volume} {96}},\
  \bibinfo {pages} {022330} (\bibinfo {year} {2017})}\BibitemShut {NoStop}%
\bibitem [{\citenamefont {Zhu}\ \emph {et~al.}(1997)\citenamefont {Zhu},
  \citenamefont {Byrd}, \citenamefont {Lu},\ and\ \citenamefont
  {Nocedal}}]{Zhu1997}%
  \BibitemOpen
  \bibfield  {author} {\bibinfo {author} {\bibfnamefont {C.}~\bibnamefont
  {Zhu}}, \bibinfo {author} {\bibfnamefont {R.~H.}\ \bibnamefont {Byrd}},
  \bibinfo {author} {\bibfnamefont {P.}~\bibnamefont {Lu}},\ and\ \bibinfo
  {author} {\bibfnamefont {J.}~\bibnamefont {Nocedal}},\ }\bibfield  {title}
  {\bibinfo {title} {Algorithm 778: {L-BFGS-B}: Fortran subroutines for
  large-scale bound-constrained optimization},\ }\href
  {https://doi.org/10.1145/279232.279236} {\bibfield  {journal} {\bibinfo
  {journal} {ACM Transactions on mathematical software (TOMS)}\ }\textbf
  {\bibinfo {volume} {23}},\ \bibinfo {pages} {550–560} (\bibinfo {year}
  {1997})}\BibitemShut {NoStop}%
\bibitem [{\citenamefont {Nielsen}(2002)}]{NielsenFidelity2002}%
  \BibitemOpen
  \bibfield  {author} {\bibinfo {author} {\bibfnamefont {M.~A.}\ \bibnamefont
  {Nielsen}},\ }\bibfield  {title} {\bibinfo {title} {A simple formula for the
  average gate fidelity of a quantum dynamical operation},\ }\href
  {https://doi.org/10.1016/S0375-9601(02)01272-0} {\bibfield  {journal}
  {\bibinfo  {journal} {Physics Letters A}\ }\textbf {\bibinfo {volume}
  {303}},\ \bibinfo {pages} {249} (\bibinfo {year} {2002})}\BibitemShut
  {NoStop}%
\bibitem [{\citenamefont {Motzoi}\ \emph {et~al.}(2009)\citenamefont {Motzoi},
  \citenamefont {Gambetta}, \citenamefont {Rebentrost},\ and\ \citenamefont
  {Wilhelm}}]{Motzoi2009}%
  \BibitemOpen
  \bibfield  {author} {\bibinfo {author} {\bibfnamefont {F.}~\bibnamefont
  {Motzoi}}, \bibinfo {author} {\bibfnamefont {J.~M.}\ \bibnamefont
  {Gambetta}}, \bibinfo {author} {\bibfnamefont {P.}~\bibnamefont
  {Rebentrost}},\ and\ \bibinfo {author} {\bibfnamefont {F.~K.}\ \bibnamefont
  {Wilhelm}},\ }\bibfield  {title} {\bibinfo {title} {Simple pulses for
  elimination of leakage in weakly nonlinear qubits},\ }\href
  {https://doi.org/10.1103/PhysRevLett.103.110501} {\bibfield  {journal}
  {\bibinfo  {journal} {Phys. Rev. Lett.}\ }\textbf {\bibinfo {volume} {103}},\
  \bibinfo {pages} {110501} (\bibinfo {year} {2009})}\BibitemShut {NoStop}%
\bibitem [{\citenamefont {Werninghaus}\ \emph {et~al.}(2021)\citenamefont
  {Werninghaus}, \citenamefont {Egger}, \citenamefont {Roy}, \citenamefont
  {Machnes}, \citenamefont {Wilhelm},\ and\ \citenamefont
  {Filipp}}]{Werninghaus2021}%
  \BibitemOpen
  \bibfield  {author} {\bibinfo {author} {\bibfnamefont {M.}~\bibnamefont
  {Werninghaus}}, \bibinfo {author} {\bibfnamefont {D.~J.}\ \bibnamefont
  {Egger}}, \bibinfo {author} {\bibfnamefont {F.}~\bibnamefont {Roy}}, \bibinfo
  {author} {\bibfnamefont {S.}~\bibnamefont {Machnes}}, \bibinfo {author}
  {\bibfnamefont {F.~K.}\ \bibnamefont {Wilhelm}},\ and\ \bibinfo {author}
  {\bibfnamefont {S.}~\bibnamefont {Filipp}},\ }\bibfield  {title} {\bibinfo
  {title} {Leakage reduction in fast superconducting qubit gates via optimal
  control},\ }\href {https://doi.org/10.1038/s41534-020-00346-2} {\bibfield
  {journal} {\bibinfo  {journal} {npj Quantum Information}\ }\textbf {\bibinfo
  {volume} {7}},\ \bibinfo {pages} {1} (\bibinfo {year} {2021})}\BibitemShut
  {NoStop}%
\bibitem [{\citenamefont {Yan}\ \emph {et~al.}(2020)\citenamefont {Yan},
  \citenamefont {Sung}, \citenamefont {Krantz}, \citenamefont {Kamal},
  \citenamefont {Kim}, \citenamefont {Yoder}, \citenamefont {Orlando},
  \citenamefont {Gustavsson},\ and\ \citenamefont {Oliver}}]{Yan2020}%
  \BibitemOpen
  \bibfield  {author} {\bibinfo {author} {\bibfnamefont {F.}~\bibnamefont
  {Yan}}, \bibinfo {author} {\bibfnamefont {Y.}~\bibnamefont {Sung}}, \bibinfo
  {author} {\bibfnamefont {P.}~\bibnamefont {Krantz}}, \bibinfo {author}
  {\bibfnamefont {A.}~\bibnamefont {Kamal}}, \bibinfo {author} {\bibfnamefont
  {D.~K.}\ \bibnamefont {Kim}}, \bibinfo {author} {\bibfnamefont {J.~L.}\
  \bibnamefont {Yoder}}, \bibinfo {author} {\bibfnamefont {T.~P.}\ \bibnamefont
  {Orlando}}, \bibinfo {author} {\bibfnamefont {S.}~\bibnamefont
  {Gustavsson}},\ and\ \bibinfo {author} {\bibfnamefont {W.~D.}\ \bibnamefont
  {Oliver}},\ }\bibfield  {title} {\bibinfo {title} {Engineering framework for
  optimizing superconducting qubit designs},\ }\href
  {https://arxiv.org/abs/2006.04130} {\bibfield  {journal} {\bibinfo  {journal}
  {arXiv:2006.04130}\ } (\bibinfo {year} {2020})}\BibitemShut {NoStop}%
\bibitem [{\citenamefont {Herrmann}\ \emph {et~al.}(2021)\citenamefont
  {Herrmann}, \citenamefont {Llima}, \citenamefont {Remm}, \citenamefont
  {Zapletal}, \citenamefont {McMahon}, \citenamefont {Scarato}, \citenamefont
  {Swiadek}, \citenamefont {Andersen}, \citenamefont {Hellings}, \citenamefont
  {Krinner} \emph {et~al.}}]{Herrmann2021}%
  \BibitemOpen
  \bibfield  {author} {\bibinfo {author} {\bibfnamefont {J.}~\bibnamefont
  {Herrmann}}, \bibinfo {author} {\bibfnamefont {S.~M.}\ \bibnamefont {Llima}},
  \bibinfo {author} {\bibfnamefont {A.}~\bibnamefont {Remm}}, \bibinfo {author}
  {\bibfnamefont {P.}~\bibnamefont {Zapletal}}, \bibinfo {author}
  {\bibfnamefont {N.~A.}\ \bibnamefont {McMahon}}, \bibinfo {author}
  {\bibfnamefont {C.}~\bibnamefont {Scarato}}, \bibinfo {author} {\bibfnamefont
  {F.}~\bibnamefont {Swiadek}}, \bibinfo {author} {\bibfnamefont {C.~K.}\
  \bibnamefont {Andersen}}, \bibinfo {author} {\bibfnamefont {C.}~\bibnamefont
  {Hellings}}, \bibinfo {author} {\bibfnamefont {S.}~\bibnamefont {Krinner}},
  \emph {et~al.},\ }\bibfield  {title} {\bibinfo {title} {Realizing quantum
  convolutional neural networks on a superconducting quantum processor to
  recognize quantum phases},\ }\href {https://arxiv.org/abs/2109.05909}
  {\bibfield  {journal} {\bibinfo  {journal} {arXiv:2109.05909}\ } (\bibinfo
  {year} {2021})}\BibitemShut {NoStop}%
\end{thebibliography}%


\begin{thebibliography}{13}%
\makeatletter
\providecommand \@ifxundefined [1]{%
 \@ifx{#1\undefined}
}%
\providecommand \@ifnum [1]{%
 \ifnum #1\expandafter \@firstoftwo
 \else \expandafter \@secondoftwo
 \fi
}%
\providecommand \@ifx [1]{%
 \ifx #1\expandafter \@firstoftwo
 \else \expandafter \@secondoftwo
 \fi
}%
\providecommand \natexlab [1]{#1}%
\providecommand \enquote  [1]{``#1''}%
\providecommand \bibnamefont  [1]{#1}%
\providecommand \bibfnamefont [1]{#1}%
\providecommand \citenamefont [1]{#1}%
\providecommand \href@noop [0]{\@secondoftwo}%
\providecommand \href [0]{\begingroup \@sanitize@url \@href}%
\providecommand \@href[1]{\@@startlink{#1}\@@href}%
\providecommand \@@href[1]{\endgroup#1\@@endlink}%
\providecommand \@sanitize@url [0]{\catcode `\\12\catcode `\$12\catcode
  `\&12\catcode `\#12\catcode `\^12\catcode `\_12\catcode `\%12\relax}%
\providecommand \@@startlink[1]{}%
\providecommand \@@endlink[0]{}%
\providecommand \url  [0]{\begingroup\@sanitize@url \@url }%
\providecommand \@url [1]{\endgroup\@href {#1}{\urlprefix }}%
\providecommand \urlprefix  [0]{URL }%
\providecommand \Eprint [0]{\href }%
\providecommand \doibase [0]{https://doi.org/}%
\providecommand \selectlanguage [0]{\@gobble}%
\providecommand \bibinfo  [0]{\@secondoftwo}%
\providecommand \bibfield  [0]{\@secondoftwo}%
\providecommand \translation [1]{[#1]}%
\providecommand \BibitemOpen [0]{}%
\providecommand \bibitemStop [0]{}%
\providecommand \bibitemNoStop [0]{.\EOS\space}%
\providecommand \EOS [0]{\spacefactor3000\relax}%
\providecommand \BibitemShut  [1]{\csname bibitem#1\endcsname}%
\let\auto@bib@innerbib\@empty
%</preamble>
\bibitem [{\citenamefont {Genest}\ \emph {et~al.}(2016)\citenamefont {Genest},
  \citenamefont {Vinet},\ and\ \citenamefont {Zhedanov}}]{S_Genest2016}%
  \BibitemOpen
  \bibfield  {author} {\bibinfo {author} {\bibfnamefont {V.~X.}\ \bibnamefont
  {Genest}}, \bibinfo {author} {\bibfnamefont {L.}~\bibnamefont {Vinet}},\ and\
  \bibinfo {author} {\bibfnamefont {A.}~\bibnamefont {Zhedanov}},\ }\bibfield
  {title} {\bibinfo {title} {Quantum spin chains with fractional revival},\
  }\href {https://doi.org/10.1016/j.aop.2016.05.009} {\bibfield  {journal}
  {\bibinfo  {journal} {Annals of Physics}\ }\textbf {\bibinfo {volume}
  {371}},\ \bibinfo {pages} {348} (\bibinfo {year} {2016})}\BibitemShut
  {NoStop}%
\bibitem [{\citenamefont {Vinet}\ and\ \citenamefont
  {Zhedanov}(2012)}]{S_Vinet2012_Krawtchouk}%
  \BibitemOpen
  \bibfield  {author} {\bibinfo {author} {\bibfnamefont {L.}~\bibnamefont
  {Vinet}}\ and\ \bibinfo {author} {\bibfnamefont {A.}~\bibnamefont
  {Zhedanov}},\ }\bibfield  {title} {\bibinfo {title} {Para-krawtchouk
  polynomials on a bi-lattice and a quantum spin chain with perfect state
  transfer},\ }\href {https://doi.org/10.1088/1751-8113/45/26/265304}
  {\bibfield  {journal} {\bibinfo  {journal} {Journal of Physics A:
  Mathematical and Theoretical}\ }\textbf {\bibinfo {volume} {45}},\ \bibinfo
  {pages} {265304} (\bibinfo {year} {2012})}\BibitemShut {NoStop}%
\bibitem [{\citenamefont {Terhal}\ and\ \citenamefont
  {DiVincenzo}(2002)}]{S_Terhal2002}%
  \BibitemOpen
  \bibfield  {author} {\bibinfo {author} {\bibfnamefont {B.~M.}\ \bibnamefont
  {Terhal}}\ and\ \bibinfo {author} {\bibfnamefont {D.~P.}\ \bibnamefont
  {DiVincenzo}},\ }\bibfield  {title} {\bibinfo {title} {Classical simulation
  of noninteracting-fermion quantum circuits},\ }\href
  {https://doi.org/10.1103/PhysRevA.65.032325} {\bibfield  {journal} {\bibinfo
  {journal} {Phys. Rev. A}\ }\textbf {\bibinfo {volume} {65}},\ \bibinfo
  {pages} {032325} (\bibinfo {year} {2002})}\BibitemShut {NoStop}%
\bibitem [{\citenamefont {Kay}(2010)}]{S_Kay2010}%
  \BibitemOpen
  \bibfield  {author} {\bibinfo {author} {\bibfnamefont {A.}~\bibnamefont
  {Kay}},\ }\bibfield  {title} {\bibinfo {title} {Perfect, efficient, state
  transfer and its application as a constructive tool},\ }\href
  {https://doi.org/10.1142/S0219749910006514} {\bibfield  {journal} {\bibinfo
  {journal} {International Journal of Quantum Information}\ }\textbf {\bibinfo
  {volume} {8}},\ \bibinfo {pages} {641} (\bibinfo {year} {2010})}\BibitemShut
  {NoStop}%
\bibitem [{\citenamefont {Cappellaro}\ \emph {et~al.}(2011)\citenamefont
  {Cappellaro}, \citenamefont {Viola},\ and\ \citenamefont
  {Ramanathan}}]{S_Cappellaro2011}%
  \BibitemOpen
  \bibfield  {author} {\bibinfo {author} {\bibfnamefont {P.}~\bibnamefont
  {Cappellaro}}, \bibinfo {author} {\bibfnamefont {L.}~\bibnamefont {Viola}},\
  and\ \bibinfo {author} {\bibfnamefont {C.}~\bibnamefont {Ramanathan}},\
  }\bibfield  {title} {\bibinfo {title} {Coherent-state transfer via highly
  mixed quantum spin chains},\ }\href
  {https://doi.org/10.1103/PhysRevA.83.032304} {\bibfield  {journal} {\bibinfo
  {journal} {Phys. Rev. A}\ }\textbf {\bibinfo {volume} {83}},\ \bibinfo
  {pages} {032304} (\bibinfo {year} {2011})}\BibitemShut {NoStop}%
\bibitem [{\citenamefont {Cade}\ \emph {et~al.}(2020)\citenamefont {Cade},
  \citenamefont {Mineh}, \citenamefont {Montanaro},\ and\ \citenamefont
  {Stanisic}}]{S_Cade2020}%
  \BibitemOpen
  \bibfield  {author} {\bibinfo {author} {\bibfnamefont {C.}~\bibnamefont
  {Cade}}, \bibinfo {author} {\bibfnamefont {L.}~\bibnamefont {Mineh}},
  \bibinfo {author} {\bibfnamefont {A.}~\bibnamefont {Montanaro}},\ and\
  \bibinfo {author} {\bibfnamefont {S.}~\bibnamefont {Stanisic}},\ }\bibfield
  {title} {\bibinfo {title} {Strategies for solving the fermi-hubbard model on
  near-term quantum computers},\ }\href
  {https://doi.org/10.1103/PhysRevB.102.235122} {\bibfield  {journal} {\bibinfo
   {journal} {Phys. Rev. B}\ }\textbf {\bibinfo {volume} {102}},\ \bibinfo
  {pages} {235122} (\bibinfo {year} {2020})}\BibitemShut {NoStop}%
\bibitem [{\citenamefont {Wittler}\ \emph {et~al.}(2021)\citenamefont
  {Wittler}, \citenamefont {Roy}, \citenamefont {Pack}, \citenamefont
  {Werninghaus}, \citenamefont {Roy}, \citenamefont {Egger}, \citenamefont
  {Filipp}, \citenamefont {Wilhelm},\ and\ \citenamefont
  {Machnes}}]{S_Wittler2021}%
  \BibitemOpen
  \bibfield  {author} {\bibinfo {author} {\bibfnamefont {N.}~\bibnamefont
  {Wittler}}, \bibinfo {author} {\bibfnamefont {F.}~\bibnamefont {Roy}},
  \bibinfo {author} {\bibfnamefont {K.}~\bibnamefont {Pack}}, \bibinfo {author}
  {\bibfnamefont {M.}~\bibnamefont {Werninghaus}}, \bibinfo {author}
  {\bibfnamefont {A.~S.}\ \bibnamefont {Roy}}, \bibinfo {author} {\bibfnamefont
  {D.~J.}\ \bibnamefont {Egger}}, \bibinfo {author} {\bibfnamefont
  {S.}~\bibnamefont {Filipp}}, \bibinfo {author} {\bibfnamefont {F.~K.}\
  \bibnamefont {Wilhelm}},\ and\ \bibinfo {author} {\bibfnamefont
  {S.}~\bibnamefont {Machnes}},\ }\bibfield  {title} {\bibinfo {title}
  {Integrated tool set for control, calibration, and characterization of
  quantum devices applied to superconducting qubits},\ }\href
  {https://doi.org/10.1103/PhysRevApplied.15.034080} {\bibfield  {journal}
  {\bibinfo  {journal} {Phys. Rev. Applied}\ }\textbf {\bibinfo {volume}
  {15}},\ \bibinfo {pages} {034080} (\bibinfo {year} {2021})}\BibitemShut
  {NoStop}%
\bibitem [{\citenamefont {Koch}\ \emph {et~al.}(2007)\citenamefont {Koch},
  \citenamefont {Terri}, \citenamefont {Gambetta}, \citenamefont {Houck},
  \citenamefont {Schuster}, \citenamefont {Majer}, \citenamefont {Blais},
  \citenamefont {Devoret}, \citenamefont {Girvin},\ and\ \citenamefont
  {Schoelkopf}}]{S_Koch2007}%
  \BibitemOpen
  \bibfield  {author} {\bibinfo {author} {\bibfnamefont {J.}~\bibnamefont
  {Koch}}, \bibinfo {author} {\bibfnamefont {M.~Y.}\ \bibnamefont {Terri}},
  \bibinfo {author} {\bibfnamefont {J.}~\bibnamefont {Gambetta}}, \bibinfo
  {author} {\bibfnamefont {A.~A.}\ \bibnamefont {Houck}}, \bibinfo {author}
  {\bibfnamefont {D.~I.}\ \bibnamefont {Schuster}}, \bibinfo {author}
  {\bibfnamefont {J.}~\bibnamefont {Majer}}, \bibinfo {author} {\bibfnamefont
  {A.}~\bibnamefont {Blais}}, \bibinfo {author} {\bibfnamefont {M.~H.}\
  \bibnamefont {Devoret}}, \bibinfo {author} {\bibfnamefont {S.~M.}\
  \bibnamefont {Girvin}},\ and\ \bibinfo {author} {\bibfnamefont {R.~J.}\
  \bibnamefont {Schoelkopf}},\ }\bibfield  {title} {\bibinfo {title}
  {Charge-insensitive qubit design derived from the cooper pair box},\ }\href
  {https://doi.org/10.1103/PhysRevA.76.042319} {\bibfield  {journal} {\bibinfo
  {journal} {Phys. Rev. A}\ }\textbf {\bibinfo {volume} {76}},\ \bibinfo
  {pages} {042319} (\bibinfo {year} {2007})}\BibitemShut {NoStop}%
\bibitem [{\citenamefont {McKay}\ \emph {et~al.}(2016)\citenamefont {McKay},
  \citenamefont {Filipp}, \citenamefont {Mezzacapo}, \citenamefont {Magesan},
  \citenamefont {Chow},\ and\ \citenamefont {Gambetta}}]{S_Mckay2016}%
  \BibitemOpen
  \bibfield  {author} {\bibinfo {author} {\bibfnamefont {D.~C.}\ \bibnamefont
  {McKay}}, \bibinfo {author} {\bibfnamefont {S.}~\bibnamefont {Filipp}},
  \bibinfo {author} {\bibfnamefont {A.}~\bibnamefont {Mezzacapo}}, \bibinfo
  {author} {\bibfnamefont {E.}~\bibnamefont {Magesan}}, \bibinfo {author}
  {\bibfnamefont {J.~M.}\ \bibnamefont {Chow}},\ and\ \bibinfo {author}
  {\bibfnamefont {J.~M.}\ \bibnamefont {Gambetta}},\ }\bibfield  {title}
  {\bibinfo {title} {Universal gate for fixed-frequency qubits via a tunable
  bus},\ }\href {https://doi.org/10.1103/PhysRevApplied.6.064007} {\bibfield
  {journal} {\bibinfo  {journal} {Phys. Rev. Applied}\ }\textbf {\bibinfo
  {volume} {6}},\ \bibinfo {pages} {064007} (\bibinfo {year}
  {2016})}\BibitemShut {NoStop}%
\bibitem [{\citenamefont {McKay}\ \emph {et~al.}(2017)\citenamefont {McKay},
  \citenamefont {Wood}, \citenamefont {Sheldon}, \citenamefont {Chow},\ and\
  \citenamefont {Gambetta}}]{S_Mckay2017}%
  \BibitemOpen
  \bibfield  {author} {\bibinfo {author} {\bibfnamefont {D.~C.}\ \bibnamefont
  {McKay}}, \bibinfo {author} {\bibfnamefont {C.~J.}\ \bibnamefont {Wood}},
  \bibinfo {author} {\bibfnamefont {S.}~\bibnamefont {Sheldon}}, \bibinfo
  {author} {\bibfnamefont {J.~M.}\ \bibnamefont {Chow}},\ and\ \bibinfo
  {author} {\bibfnamefont {J.~M.}\ \bibnamefont {Gambetta}},\ }\bibfield
  {title} {\bibinfo {title} {Efficient {Z} gates for quantum computing},\
  }\href {https://doi.org/10.1103/PhysRevA.96.022330} {\bibfield  {journal}
  {\bibinfo  {journal} {Phys. Rev. A}\ }\textbf {\bibinfo {volume} {96}},\
  \bibinfo {pages} {022330} (\bibinfo {year} {2017})}\BibitemShut {NoStop}%
\bibitem [{\citenamefont {Zhu}\ \emph {et~al.}(1997)\citenamefont {Zhu},
  \citenamefont {Byrd}, \citenamefont {Lu},\ and\ \citenamefont
  {Nocedal}}]{S_Zhu1997}%
  \BibitemOpen
  \bibfield  {author} {\bibinfo {author} {\bibfnamefont {C.}~\bibnamefont
  {Zhu}}, \bibinfo {author} {\bibfnamefont {R.~H.}\ \bibnamefont {Byrd}},
  \bibinfo {author} {\bibfnamefont {P.}~\bibnamefont {Lu}},\ and\ \bibinfo
  {author} {\bibfnamefont {J.}~\bibnamefont {Nocedal}},\ }\bibfield  {title}
  {\bibinfo {title} {Algorithm 778: {L-BFGS-B}: Fortran subroutines for
  large-scale bound-constrained optimization},\ }\href
  {https://doi.org/10.1145/279232.279236} {\bibfield  {journal} {\bibinfo
  {journal} {ACM Transactions on mathematical software (TOMS)}\ }\textbf
  {\bibinfo {volume} {23}},\ \bibinfo {pages} {550–560} (\bibinfo {year}
  {1997})}\BibitemShut {NoStop}%
\bibitem [{\citenamefont {Sete}\ \emph {et~al.}(2021)\citenamefont {Sete},
  \citenamefont {Didier}, \citenamefont {Chen}, \citenamefont {Kulshreshtha},
  \citenamefont {Manenti},\ and\ \citenamefont {Poletto}}]{S_Sete2021}%
  \BibitemOpen
  \bibfield  {author} {\bibinfo {author} {\bibfnamefont {E.~A.}\ \bibnamefont
  {Sete}}, \bibinfo {author} {\bibfnamefont {N.}~\bibnamefont {Didier}},
  \bibinfo {author} {\bibfnamefont {A.~Q.}\ \bibnamefont {Chen}}, \bibinfo
  {author} {\bibfnamefont {S.}~\bibnamefont {Kulshreshtha}}, \bibinfo {author}
  {\bibfnamefont {R.}~\bibnamefont {Manenti}},\ and\ \bibinfo {author}
  {\bibfnamefont {S.}~\bibnamefont {Poletto}},\ }\bibfield  {title} {\bibinfo
  {title} {Parametric-resonance entangling gates with a tunable coupler},\
  }\href {https://doi.org/10.1103/PhysRevApplied.16.024050} {\bibfield
  {journal} {\bibinfo  {journal} {Phys. Rev. Applied}\ }\textbf {\bibinfo
  {volume} {16}},\ \bibinfo {pages} {024050} (\bibinfo {year}
  {2021})}\BibitemShut {NoStop}%
\bibitem [{\citenamefont {Burkhart}\ \emph {et~al.}(2021)\citenamefont
  {Burkhart}, \citenamefont {Teoh}, \citenamefont {Zhang}, \citenamefont
  {Axline}, \citenamefont {Frunzio}, \citenamefont {Devoret}, \citenamefont
  {Jiang}, \citenamefont {Girvin},\ and\ \citenamefont
  {Schoelkopf}}]{S_Burkhart2021}%
  \BibitemOpen
  \bibfield  {author} {\bibinfo {author} {\bibfnamefont {L.~D.}\ \bibnamefont
  {Burkhart}}, \bibinfo {author} {\bibfnamefont {J.~D.}\ \bibnamefont {Teoh}},
  \bibinfo {author} {\bibfnamefont {Y.}~\bibnamefont {Zhang}}, \bibinfo
  {author} {\bibfnamefont {C.~J.}\ \bibnamefont {Axline}}, \bibinfo {author}
  {\bibfnamefont {L.}~\bibnamefont {Frunzio}}, \bibinfo {author} {\bibfnamefont
  {M.}~\bibnamefont {Devoret}}, \bibinfo {author} {\bibfnamefont
  {L.}~\bibnamefont {Jiang}}, \bibinfo {author} {\bibfnamefont
  {S.}~\bibnamefont {Girvin}},\ and\ \bibinfo {author} {\bibfnamefont
  {R.}~\bibnamefont {Schoelkopf}},\ }\bibfield  {title} {\bibinfo {title}
  {Error-detected state transfer and entanglement in a superconducting quantum
  network},\ }\href {DOI:https://doi.org/10.1103/PRXQuantum.2.030321}
  {\bibfield  {journal} {\bibinfo  {journal} {PRX Quantum}\ }\textbf {\bibinfo
  {volume} {2}},\ \bibinfo {pages} {030321} (\bibinfo {year}
  {2021})}\BibitemShut {NoStop}%
\end{thebibliography}%
